\documentstyle[preprint,aps,pre,eqsecnum]{revtex}  
\renewcommand \baselinestretch{1.5}
\draft
\input epsf
\tightenlines
\begin{document}
\preprint{}
\title{Casimir Forces between Spherical Particles\vspace{2mm}
in a Critical Fluid and Conformal Invariance}\vspace{25mm}
\author{E. Eisenriegler$^1$ and U. Ritschel$^{2}$}
\vspace{10mm}
\address{$^1$Institut f\"ur Festk\"orperforschung, Forschungszentrum
J\"ulich,
52425 J\"ulich (F\,R\ Germany)\\
$^2$Fachbereich Physik, Universit\"at GH Essen, 45117 Essen
(F\,R\ Germany)}

\maketitle
\narrowtext
\vspace{16mm}
\renewcommand \baselinestretch{1.1}
\begin{abstract} Mesoscopic particles immersed in a critical fluid
experience long-range Casimir forces due to critical
fluctuations. Using field theoretical methods,
we investigate the Casimir interaction between two
spherical particles and between a single particle and a planar
boundary of the fluid. We exploit the conformal symmetry at the
critical point to map both cases onto a highly symmetric geometry
where the fluid is bounded by two {\it concentric} spheres with radii
$R_-$ and $R_+$. In this geometry
the singular part of the free energy ${\cal F}$
only depends upon the ratio
$R_-/R_+$, and the stress tensor, which we use to calculate ${\cal
F}$, has
a particularly
simple form. Different boundary conditions
(surface universality classes) are considered, which either break or
preserve the order-parameter symmetry. We also
consider profiles of thermodynamic densities
in the presence of two spheres. Explicit results are presented for an
ordinary critical point to leading order in $\epsilon=4-d$
and, in the case of preserved symmetry, for the Gaussian
model in arbitrary spatial dimension $d$. Fundamental short-distance
properties, such as profile behavior near a
surface or the behavior if a sphere has a `small' radius,
are discussed and verified. The relevance for
colloidal solutions is pointed out.
\end{abstract}

\renewcommand \baselinestretch{1.45}

\setcounter{equation}{0}
\section{Introduction}\label{1}
Consider a fluid system close to a critical point, for
example a fluid near the
liquid-gas critical point, a binary liquid
near the consolute point, or liquid $^4{\rm He}$ near the
$\lambda$--point. Due to the long correlation
length a local perturbation, as provided by the wall of the container
or by the surfaces of immersed colloid particles, is not screened
within a few atomic layers but influences the system over appreciable
distances.
This gives rise to a long-range
interaction between immersed particles or a particle and a wall, as
pointed out by Fisher and de Gennes \cite{FideG78,deG80}.
The forces may be termed critical Casimir forces
\cite{Krech94}
on account of the close formal relationship with the usual
Casimir forces \cite{Casimir48} due to quantum
fluctuations of the electromagnetic field.
An interesting feature of the critical Casimir effect is that it can be
turned
on and off by changing the magnitude of the correlation length, e.g.
by changing the temperature.

The simplest and most-considered geometry for the critical
(as well as the electromagnetic) Casimir effect is the parallel-plate
geometry.
Results have been obtained in spatial dimensions $d=2$
by using conformal invariance
\cite{Bloete86,Affleck,Cardy86,BuXu91,B+E94} and
in $d=4-\epsilon$ \cite{Symanzik81,InNi86,Krech91,EiSt}.
In this paper we consider
{\it spherical boundary surfaces}, i.e.
spheres immersed in a critical fluid.
Many of the general model-independent properties for this
geometry are also discussed in a recent letter by Burkhardt
and one of the present authors \cite{B+E95}.

The spherical shape is of immediate relevance for
colloidal
particles \cite{Beysens82,Ducker91}, and besides the
planar boundary it is the simplest boundary
shape for theoretical analyses.
The reason is that systems right at the critical
point display an invariance not only with respect to
homogenous dilatations but also with respect to conformal
or angle-preserving
coordinate transformations \cite{Polyakov70}. The class of
special conformal transformations \cite{CardyRev} which exists in
arbitrary spatial dimension $d$ maps spheres onto spheres.
This has been used to study the critical
properties of an infinite system with a single spherical hole
(particle) and of a finite system with a spherical boundary
by conformally mapping results for the half--space \cite{B+E,foo1}.

An infinite system with {\it two} arbitrary non-overlapping
spherical holes (particles) may be obtained
by conformally mapping a highly
symmetric finite system which is bounded by two concentric spheres
\cite{G+R}. This is shown in Fig.\ 1 and follows from the special
conformal
transformation
\begin{equation}\label{cotr}
\frac{{\rm\bf r}^{\,\prime} }{r^{\,\prime\,2}}=
\frac{{{\rm \bf r}}+{{\rm \bf R}}}{|{{\rm \bf r}}+{{\rm \bf R}}|^2}
-\frac{{{\rm \bf R}}}{2R^2}\,
\end{equation}
which maps the sphere of radius $R$ centered at the origin of
${{\rm \bf r}}$-space onto a
plane in ${{\rm \bf r}}^{\prime}$-space that is perpendicular to
${{\rm \bf R}}$ and
passes through the origin.
Two spheres
with fixed radii $R_{-}$, $R_+$ centered about the origin of
${\rm \bf r}$-space (and depicted in Fig.\ 1\,a) map onto
two spherical particles in ${\rm \bf r}^{\prime}$-space, as shown in
Fig.\ 1\,b.
The points of intersection ${\rm I,\>II,\>III,\>IV}$
with the ${r}_{\bot}^{\prime}$-axis
have the coordinates
\begin{equation}\label{inte}
\left(r_{\bot {\rm I}}^{\prime},\,r_{\bot {\rm II}}
^{\prime},\,r_{\bot {\rm III}}^{\prime},\,r_{\bot {\rm
IV}}^{\prime}\right)=
2R\,\left(-\frac{R_+ +R }{R_+-R},\, -\frac{R_+ - R }{R_+ + R},\,
\frac{R - R_- }{R + R_-},\, \frac{R+ R_- }{R-R_-}\right)
\end{equation}
with $R_-<R<R_+$.

Two cases of particular interest will be studied
in detail below:
\begin{itemize}
\item[(i)] For $R=\sqrt{R_-R_+}$, the
{\it two spheres} in ${\rm \bf r}^{\,\prime}$-space
have {\it equal size} (SES), since ${r}_{\bot {\rm I}}^{\,\prime}=-
{r}_{\bot {\rm IV}}^{\,\prime}$ and
${r}_{\bot {\rm II}}^{\,\prime}=-{r}_{\bot {\rm III}}^{\,\prime}$.
\item[(ii)] For $R\rightarrow R_+$, the left sphere
reduces to a plane through the origin, which is denoted
by a broken line in Fig.\ 1\,b. This corresponds to
{\it one sphere}
with radius $R_1$ near a
{\it planar wall} (SPW) of the container.
\end{itemize}

One may also consider a single spherical particle immersed in
a critical fluid inside a {\it spherical
container}.
This geometry is obtained from
the concentric geometry via (\ref{cotr}) when $R$ is chosen {\it
outside} the
interval $\left(R_-,R_+\right)$. For
$R>R_+$ the order of the
intersection points is ${\rm II,\>III,\>IV,\>I}$,
as in the
concentric geometry of Fig.\ 1\,a, and sphere $a$ remains inside
sphere $b$. For $0<R<R_-$ the order is
${\rm IV,\>I,\>II,\>III}$, and sphere $a$ is located outside sphere
$b$.

An important parameter of the two-sphere geometry is the {\it
cross ratio} of intersection points, which remains invariant under
conformal
transformations \cite{Polyakov70,CardyRev}. Instead of working
directly with the cross ratio, it is
convenient to introduce the related invariant quantity
\begin{equation}\label{kapp}
\kappa=1+2\,\frac{\left({r}_{\bot {\rm IV}}^{\,\prime} -
{ r}_{\bot {\rm I}}^{\,\prime}\right)\left({ r}_{\bot {\rm
III}}^{\,\prime}-
{ r}_{\bot {\rm II}}^{\,\prime}\right)}{\left({ r}_{\bot {\rm
IV}}^{\,\prime}-
{ r}_{\bot {\rm III}}^{\,\prime}\right)\left({ r}_{\bot {\rm
II}}^{\,\prime}-
{ r}_{\bot {\rm I}}^{\,\prime}\right)}\>,
\end{equation}
which can be written as
\begin{equation}\label{kap2}
\kappa=\frac{|s_{12}^2-R_1^2-R_2^2|}{2R_1R_2}
\end{equation}
in either of the above geometries. Here $R_1$ and $R_2$ are the radii
of spheres $a$ and $b$, respectively, and $s_{12}$
is the distance between their centers.
On comparing with the concentric geometry of Fig.\ 1\,a, where
$s_{12}=0$ and
where $R_{1}$ and $R_2$ have been denoted by $R_-$ and $R_+$,
respectively, one finds from (\ref{kap2})
\begin{equation}\label{kap3}
\kappa=\frac12\left(\rho+\rho^{-1}\right)\,, \qquad1<\kappa<\infty\>,
\end{equation}
with
\begin{equation}\label{rhof}
\rho=\frac{R_-}{R_+}\>\>,\qquad 0<\rho<1\>.
\end{equation}
One can also obtain
(\ref{kap3}) by inserting (\ref{inte}) into (\ref{kapp}), which
demonstrates
the conformal invariance of $\kappa$ since the parameter $R$ of
(\ref{cotr}) drops out. It follows from (\ref{kapp}) or
(\ref{kap2}) that $\kappa=1$ corresponds to
two particles that touch, while $\kappa\rightarrow\infty$
describes the situation where one or both
of the radii $R_1,\>R_2$ are much smaller than
the remaining lengths in the two-sphere configuration.

Now let us turn our attention to the free energy.
For two non-overlapping spherical particles in an infinite
critical medium we consider
\begin{equation}\label{del1}
\delta F(s_{12},\,R_1,\,R_2\,)=F_{a,\,b}(s_{12},\,R_1,\,R_2\,)
-U_a(R_1)-U_b(R_2)\>,\quad s_{12}\ge R_1+R_2\>.
\end{equation}
Here $F$ is the singular part \cite{Privman84} of the free energy
for immersing two particles into an infinitely extended fluid,
while $U$ is that for immersing a single
particle. $\delta F$ determines the Casimir force,
$-\partial \,\delta F/\partial s_{12}$, and it vanishes as the separation
$s_{12}$ tends to infinity with $R_1,\,R_2$ fixed.
For the single particle ($a,\,R_1$) in the fluid with spherical
confinement ($b,\,R_2$) one may consider a similar expression
$\delta F(s_{12},\,R_1,\,R_2\,)$ with $s_{12}+R_1\le R_2$, where
$U_a(R_1)$ and $\bar U_b(R_2)$, the free energy of the spherical container
without a particle, are subtracted from the total free energy.
This expression vanishes as $R_2$ tends to infinity with $R_1$ and $s_{12}$
fixed. Right at the bulk critical point the quantities $\delta F$
are invariant under special conformal transformations \cite{CardyRev,foo2}
and can only depend on the conformal invariant $\kappa$ in (\ref{kap2}):
\begin{equation}\label{delf}
\delta F(s_{12},\,R_1,\,R_2\,)=k_BT_c\,{\cal F}(\kappa)
\end{equation}
It follows from the above arguments that
$\delta F$ for two particles in the infinite critical liquid and
$\delta F$ for the single particle in the spherically confined liquid
both obey (\ref{delf}) with the same
function ${\cal F}(\kappa)$. We note
two special cases. First, for a single sphere with radius $R_1$ and
with its closest point at a distance $D$ from a planar wall (SPW),
the Casimir interaction is
\begin{equation}\label{fspw}
\delta F_{\rm \scriptscriptstyle SPW}(D,R_1)=k_BT_c\;{\cal
F}\left(1+D/R_1\right)\>.
\end{equation}
This follows from (\ref{delf}) by substituting $s_{12}=R_1+R_2+D$ and
with $R_2\rightarrow \infty$. Second, for the concentric geometry
(CON)
of Fig.\ 1\,a, the
result reads
\begin{equation}\label{fcon}
\delta F_{\rm \scriptscriptstyle CON}(R_-,R_+)=k_BT_c\,{\cal
F}\left(\kappa(R_-/R_+)\right)\>,
\end{equation}
with $\kappa(\rho)$ given by (\ref{kap3}).
Due to the high symmetry, the concentric
geometry is especially convenient for the explicit evaluation
of the universal scaling function ${\cal F}(\kappa)$.

The scaling function ${\cal F}$ depends on both
the bulk universality class
\cite{WilKo} and the surface universality classes
$a,\>b$ \cite{Binder83,Diehl86} of
the critical system. For a liquid-gas critical point or a critical
consolute point of a binary mixture, $a$ and $b$ correspond to
the Ising model with surface magnetic fields
\cite{Binder83,Diehl86,BuCa87,BuDi94} either parallel
$(\uparrow ,\uparrow)$ or
antiparallel $(\uparrow ,\downarrow)$ depending on whether the two
surfaces favor the same or different signs of the order parameter.
Liquid $^4$He at the $\lambda$--point belongs to
the bulk universality class of the $XY$ model, and surfaces are
expected
to suppress the superfluid order \cite{Dohm}, corresponding to the
surface universality class $O$ of the `ordinary transition'
\cite{Binder83,Diehl86}.
One may also consider tricritical bulk universality classes
both for mixtures of ordinary liquids and of $^3$He--$^4$He. It
is also interesting to consider the multicritical
surface universality class $SB$ of the `special transition'
\cite{Binder83,Diehl86}.

Exact results for ${\cal F}$ have been obtained by
Burkhardt and Eisenriegler \cite{B+E95}
for the critical Ising model with pairs of surface universality
classes $(a,b)=
(O,O),\>(O,\uparrow),\>(\uparrow ,\uparrow)$, and
$(\uparrow ,\downarrow)$
in spatial dimension $d=2$. In the present paper we investigate
${\cal F}$ for $d$ near the upper critical dimension $d_c=4$. For the
combinations
$(a,b)=
(\uparrow ,\uparrow)$, $(\uparrow ,\downarrow)$, $(\uparrow ,O)$, and
$(\uparrow ,SB)$, where the order parameter symmetry is broken, the
leading behavior in the $\epsilon$-expansion follows
from
mean-field theory. This will be considered in Sec.\ \ref{3}. Here we
make close contact to the work of Indekeu et al. \cite{InNi86} for the
parallel-plate geometry
and of Gnutzmann and Ritschel \cite{G+R} for the order-parameter
profile
in the concentric geometry for $(\uparrow ,\uparrow)$. For the
symmetry-preserving combinations $(O,O)$, $(O,SB)$, and $(SB,SB)$, the
leading behavior is that of the Gaussian model. Since the Gaussian
model is conformally invariant for all $d>2$, it is instructive to
consider the above pairs of boundaries in general $d$.
This is carried
out in Sec.\ \ref{4}. The parallel-plate result
(Krech and Dietrich \cite{Krech91}) emerges as a special case.

The explicit evaluation
of the different universal scaling functions ${\cal F}$
in Secs.\ \ref{3} and \ref{4} are preceded in Sec.\ \ref{2} by a
discussion of general relationships involving ${\cal F}$ which hold
for all spatial dimensions and universality classes of interest. We
discuss the limits $\kappa\rightarrow 1$ and
$\kappa\rightarrow\infty$, which have also been addressed in
Ref. \cite{B+E95} and correspond to $\rho\rightarrow 1$ and $\rho
\rightarrow 0$, respectively. For $\rho\rightarrow 1$, the scaling
functions ${\cal F}$
are determined by the amplitudes of the Casimir effect in {\it
parallel-plate}
geometry
\cite
{Bloete86,Affleck,Cardy86,BuXu91,B+E94,Symanzik81,InNi86,Krech91,EiSt}.
For $\rho\rightarrow 0$, we use the {\it small-sphere}
expansion \cite{B+E95} (a kind of short-distance expansion)
to relate ${\cal F}$ to the amplitudes of the bulk correlation
function and of the profile in the half--space of either
the order parameter or the energy density,
depending on the pair $a,b$ of
universality classes. We also discuss how ${\cal
F}$ can be conveniently calculated from
the thermal average of the {\it stress tensor}
\cite{CardyRev}. Furthermore, we
discuss the profiles of thermodynamic densities in the presence of two
spheres \cite{G+R}. For the particular case of a sphere and a planar
wall (SPW), we use a short-distance
expansion \cite{Cardy90,EKD,EiSt} to relate the
behavior of the
profile near the wall to the average of the stress tensor at the wall,
i.e. to ${\cal F}$. All the
general relationships are consistent with the explicit results
in Secs.\ \ref{3} and \ref{4}.

Sec.\ \ref{5} contains our conclusions.

\setcounter{equation}{0}

\section{General relationships}\label{2}
Here we discuss several important relationships which involve the
scaling function ${\cal F}$ of the free energy. They are of a general
character and apply to most of the bulk and surface universality
classes and spatial dimensions $d$ which are of interest.
\subsection{Free energy and the stress tensor}\label{2A}
While conformal coordinate transformations of the two-sphere geometry
preserve
$\kappa$ (and thus leave $\delta F$ invariant), certain
non-conformal transformations preserve the spherical shape
of the boundaries but change $\kappa$.
In equation (\ref{aeq1}) we consider a
transformation which increases the distance
$s_{12}$ of the two non-overlapping spheres of
Fig.\ 1\,b keeping $R_1$ and $R_2$ fixed. This is
directly related to the Casimir force.
In the concentric geometry
of Fig.\ 1\,a one may apply the infinitesimal transformation ${\rm \bf
r}\rightarrow
\,\widehat{{\rm \bf r}}$ defined by
\begin{equation}\label{thet}
\widehat{{\rm \bf r}}-{\rm \bf r}\>=\>\alpha\;{\rm \bf r}
\>\Theta(r-r_0)\>\>,
\qquad \alpha\ll 1
\end{equation}
which increases the radial component of ${\;\rm \bf r\;}$ by a fraction
$\alpha$ if \ $r>r_0$. Choosing $r_0$ between $R_-$ and
$R_+$, one sees that $R_+$ increases while
$R_-$ remains unchanged.  Consequently, $\rho$ and $\kappa$ are
changed. The effect of these transformations on $\delta F$ can be
written as an integral over the stress-energy tensor $T_{kl}$
\cite{CardyRev}. In the concentric case
\begin{equation}\label{deef}
\rho\;\frac{\rm d}{{\rm d}\rho}\;\delta F_{{\rm \scriptscriptstyle
CON}}(\rho)=k_BT_c\>S_d\;r_0^d\>\langle T_{nn}(r_0)
\rangle_{{\rm\scriptscriptstyle CON}}\>,
\end{equation}
as we show in more detail in Appendix A. Here
$S_d=2\pi^{d/2}/\Gamma(d/2)$ is the surface area
of the $d$-dimensional
unit sphere, and $T_{nn}(r_0)$ is the radial component (normal to the
spherical surfaces) of the stress tensor at a
distance $r_0$ from the center. The thermal average at the
right hand side of
Eq.\ (\ref{deef}) is taken at the critical
fixed point, where the trace
of the stress tensor vanishes \cite{CardyRev}.
Since $r_0$ is arbitrary in
the interval $(R_-,R_+)$,
\begin{equation}\label{tenn}
\langle T_{nn}(r_0)\rangle_{{\rm\scriptscriptstyle
CON}}=(2r_0)^{-d}\,Y(\rho)\>\>,
\end{equation}
where we have included a factor $2^{-d}$ for convenience
below. In evaluating
${\cal F}(\kappa)$ and the Casimir forces
in Sec.\ \ref{3}
and \ref{4}, we have found it convenient to first calculate the
scaling function $Y$ defined in (\ref{tenn}) and then integrate the
result according
to (\ref{deef}).

\subsection{Free energy for ${\bf \kappa\rightarrow 1}$}\label{2B}
The limit $\rho\rightarrow 1$ corresponds to the parallel-plate
geometry. The concentric geometry becomes
equivalent to the parallel-plate geometry if we let
$R_+\rightarrow\infty$ while keeping the width $L=R_+-R_-$ fixed.
In
this limit $1-\rho=L/R_+$ tends to zero. With the usual
notation \cite{Krech94,InNi86,Krech91} $\Delta_{a,b}/L^{d-1}$ for the
Casimir free energy per $k_BT$ and surface area of one of the two
parallel plates $a,b$,
\begin{mathletters}\label{lim}
\begin{equation}\label{limf}
\delta F_{{\rm\scriptscriptstyle CON}}(\rho)\rightarrow
k_BT_c\;\Delta_{a,b}\;(1-\rho)^{1-d} \,S_d
\end{equation}
and
\begin{equation}\label{limy}
Y(\rho)\rightarrow \Delta_{a,b}\,(d-1)\,2^d\,(1-\rho)^{-d}
\end{equation}
\end{mathletters}\noindent
\noindent
as $\rho\rightarrow 1$.
Eq.\,(\ref{limf}) also determines the Casimir interaction of
non-overlapping spheres that nearly touch. This follows from the form
$S_d \Delta_{a,b}
\left[2(\kappa-1)\right]^{-(d-1)/2}$ of ${\cal F}$ for
$0<\kappa-1\ll 1$ as implied by (\ref{fcon}) and (\ref{kap3}).
In this geometry the limiting form reads \cite{B+E95}
\begin{equation}\label{plaw}
\delta F(s_{12},R_1,R_2)\rightarrow\,k_BT_c\,S_d\,\Delta_{a,b}
\left[2(R_1^{-1}+R_2^{-1})\,D\right]^{-(d-1)/2}\>.
\end{equation}
Here (\ref{delf}) and (\ref{kap2}) have been used, $D\equiv
s_{12}-R_1-R_2$ is the distance between the
closest points of the two spherical surfaces
(i.e. $r_{\bot\,{\rm III}}^{\,\prime}-r_{\bot\,{\rm II}}^{\,\prime}$
in
Fig.\ 1\,b), and
we assume $D\ll R_1,R_2$. In the limit $R_2^{-1}\rightarrow 0$, the
right hand side
of (\ref{plaw}) determines the
Casimir interaction $\delta F_{\rm \scriptscriptstyle SPW}(D,R_1)$ of
a single sphere close to a planar wall $(D\ll R_1)$.

\subsection{Small-sphere expansion and free energy for
$\kappa\rightarrow
\infty$}\label{2C}
In the other limit $\rho\rightarrow 0$, where the radius of the inner
sphere is much smaller than that of the outer boundary, one may use
a short-distance expansion introduced in Ref. \cite{B+E95}. In this
expansion the Boltzmann weight $\exp(-{\cal H}_S)$ which generates in
the critical system a spherical hole $S$ with center at ${\rm \bf
r}_S$ and radius $R_S$ is written in terms of
local fluctuating quantities (operators) $\Xi({\rm \bf r}_S)$
and non-fluctuating
amplitudes ${\cal X}$ in the form
\begin{mathletters}\label{sms}
\begin{equation}\label{sms1}
\frac{{\rm e}^{-{\cal H}_S}}{\langle
{\rm e}^{-{\cal H}_S}\rangle_{\rm bulk}}=1+\sigma
\end{equation}
with
\begin{equation}\label{sms2}
\sigma=\sum_{\Xi}\,{\cal X}_{\Xi}(R_S)\;\Xi({\rm \bf r}_S)\>.
\end{equation}
\end{mathletters}\noindent
\noindent
The relation (\ref{sms}) holds in thermal averages
or correlation functions containing other fluctuating quantities
(operators) or boundaries \cite{foo3} with distances from ${\rm \bf
r}_S$ that are much
larger than $R_S$.
We mention a few examples where (\ref{sms}) is useful:
Consider the thermal average $\langle \tilde\Psi \rangle_{S\> {\rm
in\,bulk}}$ of a
{\it primary} operator $\tilde\Psi({\rm \bf r})$ \cite{foo3a}
such as the order parameter or the energy density in an infinite
critical system (bulk) with a spherical hole $S$. By means of
(\ref{sms1}) one finds
\begin{equation}\label{hole}
\langle \tilde\Psi({\rm \bf r})\rangle_{S\,{\rm in\,bulk}}=
\langle{\rm e}^{-{\cal H}_S}\,\tilde\Psi({\rm \bf r})\rangle_{\rm bulk}/
\langle{\rm e}^{-{\cal H}_S}\rangle_{\rm bulk}=
\langle(1+\sigma)\cdot \tilde\Psi({\rm \bf r})\rangle_{\rm bulk}
\end{equation}
since the bulk average of $\sigma$ vanishes.
As another example consider the concentric geometry, which may be
viewed as a critical system inside a full spherical container (fc) with
Hamiltonian
${\cal H}_{\rm fc}$, where a hole $S$ is generated by adding ${\cal
H}_S$.
We are interested in the combination of free energies
\begin{eqnarray}\label{spfc}
\lefteqn{\left(F_{S\,{\rm in\;fc}}-F_{\rm fc}\right)-
\left(F_{S\,{\rm in \; bulk}}-F_{\rm bulk}\right)=}\nonumber\\
& &\hspace{15mm}k_BT_c\>\left[
-{\rm ln}\>\langle{\rm e}^{-{\cal H}_S}\rangle_{\rm fc}
+{\rm ln}\>\langle{\rm e}^{-{\cal H}_S}\rangle_{\rm bulk}\right]
=-k_BT_c\>{\rm ln}\>\langle 1+\sigma\rangle_{\rm fc}\,,
\end{eqnarray}
the singular part of which is the function
$\delta F_{{\rm\scriptscriptstyle CON}}$ in (\ref{fcon}). As a third
example one may use (\ref{sms1}) also to study the behavior of
$\langle\tilde\Psi({\rm \bf r})\rangle$
in the concentric geometry as $R_-\rightarrow 0$. One finds
\begin{equation}\label{tcon}
\langle\tilde\Psi({\rm \bf r})\rangle_{\rm \scriptscriptstyle CON}-\langle
\tilde\Psi({\rm \bf r})\rangle_{\rm fc}\longrightarrow
\langle\sigma\;\tilde\Psi({\rm \bf r})\rangle_{\rm fc}
-\langle \sigma\rangle_{\rm fc}\;\langle\tilde\Psi({\rm \bf
r})\rangle_{\rm fc}
\end{equation}
to first order in $\sigma$. This relation also applies for
$\tilde\Psi\rightarrow T_{nn}$, (see (\ref{tenn})).
Further,
the expansion (\ref{sms2}) holds also
if the distant boundary is not
a sphere concentric to $S$ but, for example, a planar wall \cite{foozyl},
and it should also apply slightly away from the critical point
\cite{foo35}.

Which quantities $\Xi$ appear in (\ref{sms2}) depends upon the
universality class of the surface of $S$. In case of the surface
universality classes $O$ and $SB$ that do not break the $O(N)$
symmetry of the $N$-component order parameter $\roarrow  \Phi$ at the
bulk critical point, all the $\Xi$'s are also $O(N)$ invariant,
and the
leading $\Xi$ is the energy density $\roarrow\Phi^{\>2}$ \cite{foo3b}.
For symmetry-breaking surfaces we only consider the case of a
one-component order parameter $\Phi$ and the class $\uparrow$ or
$\downarrow$. In this case $\Xi$'s, which are
odd and even under $\Phi\rightarrow -\Phi$
both appear in the expansion. The
leading $\Xi$ is the order parameter $\Phi$ and the leading
even $\Xi$ is the energy density.

In (\ref{sms}) $R_S$ is assumed to be large on a microscopic
scale. The amplitudes ${\cal X}$ for
$\Xi=\Phi$ or $\Phi^2$ can be derived explicitly.
They follow from the known result \cite{B+E} for the
profile $\langle\tilde\Psi({\rm \bf r})\rangle_{S\,{\rm in\,bulk}}$ of the
primary operators $\tilde\Psi=\Phi$ or $\Phi^2$ by
comparing with (\ref{hole}). From the
leading behavior for $|r-r_S|\gg R_S$,
\begin{equation}\label{iden}
{\cal
X}_{\Psi}=\frac{A_a^{\Psi}}{B_{\Psi}}\>\left(R_S\right)^{x_{\Psi}}
\end{equation}
for $\Psi=\Phi$ or $\Psi=\Phi^2$, respectively \cite{B+E95}.
Here $a$ denotes the
surface universality class of $S$, $A$ is the amplitude in the
critical profile
\begin{equation}\label{ampl}
\langle\Psi({\rm\bf r}_{\parallel},z)\rangle^{(a)}_{\rm half\;space}=
A_a^{\Psi}\;
(2\, z)^{-x_{\Psi}}
\end{equation}
of $\Psi$ in a half--space with surface universality class $a$, and $B$
is the
amplitude of the bulk correlation function
\begin{equation}\label{corr}
\langle\Psi({\rm \bf r}\;)\Psi(0)\rangle_{\rm bulk}=
B_{\Psi}\,r^{-2x_{\Psi}}\>,
\end{equation}
which also identifies $x_{\Psi}$ as the bulk exponent of $\Psi$.

Now we may calculate $\delta F_{{\rm \scriptscriptstyle CON}}$ for
small $\rho$ as the singular part of
(\ref{spfc}) by using (\ref{sms2}) and (\ref{iden}), with the result
\begin{mathletters}\label{frc}
\begin{equation}\label{frco}
\delta F_{{\rm \scriptscriptstyle CON}}(\rho)/k_BT_c
\longrightarrow-{\cal X}_{\Psi}\cdot\langle\Psi(0)\rangle_{\rm fc} =
-\frac{A_a^{\Psi}\,A_b^{\Psi}}{B_{\Psi}}\>\rho^{x_{\Psi}}\>,
\end{equation}
where $\Psi=\Phi$ if both surfaces $a,b$ are symmetry-breaking and
$\Psi=\Phi^2$ if one or both preserve the symmetry of the order
parameter. Here we have identified $R_S$ with $R_-$ of Fig.\ 1\,a and
have used the fact that the profile $\langle\Psi(0)\rangle_{\rm fc}$
at the center of a full spherical container with radius $R_+$ and surface
universality class $b$ is given by
$A_b^{\Psi}\left(R_+\right)^{-x_{\Psi}}$ (see \cite{B+E}). Below
we will also need the small-$\rho$ limit
\begin{equation}\label{ylim}
Y(\rho)\rightarrow - 2^d\,S_d^{-1}\,x_{\Psi}\,
\rho^{x_{\Psi}}\,\frac{A_a^{\Psi}\,A_b^{\Psi}}{B_{\Psi}}
\end{equation}
\end{mathletters}\noindent
\noindent
of the stress-tensor scaling function defined in (\ref{tenn}). This
follows either from (\ref{frco}) by means of (\ref{deef}) or by
inserting
(\ref{sms2}) and (\ref{iden}) into (\ref{tcon}) with
$\tilde\Psi({\rm \bf r})\rightarrow T_{nn}(r_0)$ and using
$\langle T_{kl}\rangle_{\rm fc}=0$ and \cite{Cardy90}
\begin{equation}\label{pste}
\langle\Psi(0)\,T_{nn}(r_0)\rangle_{\rm
fc}=-r_0^{-d}\left(R_+\right)^{-x_{\Psi}}\,x_{\Psi}\,S_d^{-1}\,
A_b^{\Psi}\>.
\end{equation}

Eq.\,(\ref{frco}) also implies, via (\ref{fcon}) and (\ref{kap3}),
the form
$-\left(A_a^{\Psi}A_b^{\Psi}/B_{\Psi}\right)(2\kappa)^{-x_{\Psi}}$ for
the scaling function ${\cal F}$ in (\ref{delf}) if $\kappa\gg 1$. This
leads to the
Casimir interaction \cite{B+E95}
\begin{mathletters}\label{fli}
\begin{equation}\label{fdis}
\delta F(s_{12},R_1,R_2)\rightarrow - k_BT_c\;
\frac{A_a^{\Psi}A_b^{\Psi}}{B_{\Psi}}\left(\frac{R_1R_2}{s_{12}^2}
\right)^{x_{\Psi}}
\end{equation}
between distant spheres ($s_{12}\gg R_1,R_2$) and
\begin{equation}\label{fdsp}
\delta F_{\scriptscriptstyle \rm SPW}
(D,R_1)\rightarrow - k_BT_c\;
\frac{A_a^{\Psi}A_b^{\Psi}}{B_{\Psi}}\left(\frac{R_1}{2D}
\right)^{x_{\Psi}}
\end{equation}
\end{mathletters}\noindent
\noindent
between a planar wall and a single distant sphere
$(D\gg R_1)$. Here Eqs.\,(\ref{kap2}) and (\ref{fspw}) have been used.

\subsection{Density profiles and short-distance expansion near a
planar surface}\label{2D}
In this section
we study the profile
$\langle\Psi({{\rm \bf r}}^{\,\prime})\rangle$ as ${\rm \bf
r}^{\,\prime}$ approaches one of the boundaries of the critical
system.
The normal component \cite{foo4}
$\langle T_{\!\bot\!\bot}({\rm \bf r}^{\,\prime})\rangle$ of the
stress-tensor average is {\it regular} near the
boundaries \cite{foo5}.
For the concentric geometry, this is apparent from (\ref{tenn}), and
it holds quite generally \cite{McAOs93,foo6}. The behavior of the order
parameter and the
energy density is quite different, however, and is
described by power-law exponents. In general these are determined by
the scaling dimensions of surface operators and are
different from bulk exponents \cite{Binder83,Diehl86}. Here we
consider the particularly simple cases where $\Psi$ approaches a
{\it planar} boundary with universality class $b$ and $(\Psi,b)$ is
either $(\Phi,\uparrow)$,  $(\Phi^2,\uparrow)$, or $(\Phi^2, O)$.
Then the leading surface operator is
$T_{\!\bot\!\bot}$ \cite{BuXu91,B+E94,EiSt,Cardy90,EKD}, which leads
to a surface exponent $d$
\cite{BuCa87,BuDi94}. The corresponding
short-distance expansion about the planar surface \cite{Diehl86} has
the form \cite{Cardy90,EKD,McAOs93,EiSt}
\begin{equation}\label{shor}
\Psi\left({\rm \bf r}^{\,\prime}_{\parallel},
r^{\,\prime}_{\bot}\right)=
A_b^{\Psi}\left(2 r_{\bot}^{\,\prime}\right)^{-x_{\Psi}}
\left[1-C_{\Psi}^{(b)}\,r^{\>\prime\,d}_{\bot}
\>\lim_{r^{\,\prime}_{\bot}\rightarrow 0}\>
T_{\!\bot\!\bot}\!\left({\rm \bf r}^{\,\prime}_{\parallel},
r^{\,\prime}_{\bot}\right)
+\ldots\right]\>.
\end{equation}
The planar surface is located in ${\rm \bf r}^{\,\prime}$-space
at $r^{\,\prime}_{\bot}=0$. The factor in front of the square bracket
is the critical half--space profile as in (\ref{ampl}) for
surface universality class $b$. $C_{\Psi}^{(b)}$ is a universal
short-distance amplitude which generally depends on $b$ \cite{foo7}.
Its behavior is known in the Gaussian model with $d>2$ and in the
$\Phi^4$-theory near the upper critical dimension $d=4$\cite{EKD,EiSt}.
The ellipses inside the square brackets in
(\ref{shor}) denote contributions of higher order in
$r^{\,\prime}_{\bot}$.
The expansion (\ref{shor}) holds in thermal averages or
correlation functions where besides the boundary at
$r^{\,\prime}_{\bot}=0$ and
$\Psi$ other boundaries and operators may be involved.

We are particularly interested in the
small $r^{\,\prime}_{\bot}$ behavior of $\langle
\Psi({\rm \bf r}^{\,\prime}_{\parallel},
r^{\,\prime}_{\bot}
\rangle_{\rm \scriptscriptstyle SPW}$ in the
sphere-near-planar-wall (SPW) geometry, which has been introduced as
case (ii) just below Eq.\,(\ref{inte}) and
in Eq.\,(\ref{fspw}). For the stress tensor
\begin{mathletters}\label{tsp}
\begin{equation}\label{tspw}
\langle T_{\!\bot\!\bot}({\rm \bf
r}^{\,\prime}_{\parallel},0)\rangle_{\rm \scriptscriptstyle SPW}=
\frac{\left[D\left(D+2R_1\right)\right]^{d/2}}{\left[{\rm \bf
r}_{\parallel}^{\,\prime\,2}+D\left(D+2R_1\right)\right]^d}\>Y(\rho)
\>,
\end{equation}
with
\begin{equation}\label{rhod}
\rho=\left[D+R_1-\sqrt{D(D+2R_1)}\right]/R_1
\end{equation}
\end{mathletters}\noindent
and $D$ as defined in (\ref{fspw}). Here we have used the general
transformation formula \cite{CardyRev,McAOs93}
\begin{mathletters}\label{tra}
\begin{equation}\label{ttra}
\langle T_{\!\bot\!\bot}({\rm \bf r}^{\,\prime})\rangle=
\left[b({\rm \bf r}^{\,\prime})\right]^{-d}\>\langle
T_{nn}(r)\rangle_{\rm \scriptscriptstyle CON}
\end{equation}
with the local scale factor
\begin{equation}\label{losc}
b({\rm \bf r}^{\,\prime})\equiv\left|\frac{\partial{\rm \bf
r}}{\partial {\rm \bf r}^{\,\prime}}\right|^{-1/d}
=1+\frac{{\rm \bf R} \cdot {\rm \bf
r}^{\,\prime}}{R^2}+\frac{r^{\,\prime \,2}}{4R^2}\>.
\end{equation}
\end{mathletters}\noindent
Setting $R=R_+$ leads to the SPW geometry on the left
hand side of (\ref{ttra})
with
\begin{equation}\label{drel}
D(D+2R_1)=r^{\,\prime}_{\bot {\rm III}}\>r^{\,\prime}_{\bot {\rm
IV}}=(2R_+)^2\>,
\end{equation}
which follows from (\ref{inte}). Using the form (\ref{tenn})
for $T_{nn}(r)$ on the
right hand side of (\ref{ttra}) and letting
$r\rightarrow R_+$ so that $r^{\,\prime}_{\bot}\rightarrow 0$
leads to (\ref{tspw}). Eq.\,(\ref{rhod})
follows e.g. from (\ref{kap3}), since the
quantity $\kappa$ from (\ref{kap2}) equals $1+D/R_1$ in the
SPW case.\vspace{4mm}

We close this section by introducing some of the tools needed
in Secs.\ \ref{3} and \ref{4}.
Critical systems with surfaces may be described \cite{Diehl86} by a
configuration probability $\sim
\exp(-{\cal H})$ for the fluctuating $N$-component order parameter
$\roarrow\Phi$,
where ${\cal H}$ separates into a bulk part
\begin{equation}\label{hbul}
{\cal H}_{\rm bulk}=\int\,{\rm d}^dr\>{\cal L}\>,\qquad
{\cal
L}=\frac12\>{\bf\nabla}\roarrow\Phi\cdot{\bf\nabla}\roarrow\Phi+\frac{
u}{4!}
\left(\,\roarrow\Phi^{2}\right)^2
\end{equation}
and surface parts ${\cal H}_{\rm surf}$. While the integration in
(\ref{hbul}) is over the interior of the critical system, e.g. over
$R_-<r<R_+$ in the concentric geometry, the surface contributions
involve
integrals (of powers of $\Phi$ and derivatives thereof) over the
surfaces of the system. Since we shall only consider systems right at
the bulk critical point, we have not included a $\Phi^2$ term in
(\ref{hbul}) \cite{foo8}.

We will also need the explicit form of the stress tensor at an
arbitrary interior point of the system. It is given by
\cite{CCJ70,Brown80}
\begin{mathletters}\label{tkl}
\begin{equation}\label{tekl}
T_{kl}({\rm \bf
r})=(\partial_k\roarrow\Phi)\,(\partial_l\roarrow\Phi)-
\delta_{kl}\,{\cal L}-{\cal J}_{kl}\>,
\end{equation}
with the canonical tensor supplemented by the
improvement term
\begin{equation}\label{impr}
{\cal J}_{kl}({\rm \bf r})=\frac14\,\frac{d-2}{d-1}
\left(1+{\cal
O}(u_R^3)\right)\left[\partial_k\partial_l-\delta_{kl}\Delta
\right]\,\roarrow\Phi^{\> 2}\>.
\end{equation}
\end{mathletters}\noindent
The quantity denoted by ${\cal O}(u_R^3)$ is needed to
renormalize the stress tensor \cite{Brown80}. Since it is of third
or higher order in the renormalized coupling $u_R$
(compare Eq. (\ref{relu}) below), it does not appear
in our explicit low-order calculations.

The thermal average of the stress tensor
in the concentric geometry takes the
rotationally-invariant form
\begin{equation}\label{troi}
\langle T_{kl}({\rm \bf r})\rangle_{\rm \scriptscriptstyle CON}=r^{-d}
\left[\frac{r_kr_l}{r^2}\,\tau^{(1)}-\frac{\delta_{kl}}{d}\,\tau^{(2)}
\right]\>.
\end{equation}
Here $\tau^{(1)},\>\tau^{(2)}$ are dimensionless quantities which
only depend on the position via
$r=|{\rm \bf r}|$. A fundamental property of the stress tensor
is the continuity equation \cite{Brown80}. In the present case
\begin{eqnarray}\label{cont}
0&=&\sum_l\,\frac{\partial\langle T_{kl}({\rm \bf r})\rangle_{\rm
\scriptscriptstyle CON}}{\partial r_l}
\nonumber\\*[3mm]
&= &r_k\,r^{-d-2}\left[-\left(\tau^{(1)}-\tau^{(2)}\right)
+r\,\frac{\mbox{d}}{\mbox{d}\, r}\left(\tau^{(1)}-
\frac{1}{d}\;\tau^{(2)}\right)\right]\>,
\end{eqnarray}
which implies the important relation
\begin{equation}\label{trel}
r\,\frac{\mbox{d}}{\mbox{d}\,r}\,r^d\,\langle T_{nn}({\rm \bf
r})\rangle_{\rm \scriptscriptstyle CON}=r^d\>
\sum_l\langle T_{ll}({\rm \bf r})\rangle_{\rm \scriptscriptstyle CON}
\end{equation}
between the $r$-dependence of the radial component and the trace of the
stress tensor. Eqs.\,(\ref{cont}) and (\ref{trel}) hold for arbitrary
$u,\,d,\,r$ and for arbitrary (rotationally-invariant) conditions at the
surfaces. If the trace vanishes, as happens \cite{Brown80} if
${\cal H}_{\rm bulk}$ is at the critical fixed point of the
renormalization group (RG), Eq.\,(\ref{trel}) implies that $r^d\langle
T_{nn}\rangle_{\rm \scriptscriptstyle CON}$ is independent of $r$.
This is consistent with Eq.\ (\ref{tenn}).

To set up the RG, we use reparametrization by minimal subtraction
of poles in $\epsilon$ \cite{Amit,ZinnJust,Brown80,Diehl86}.
We note in particular the relation
\begin{equation}\label{relu}
u = 16\pi^2 \mu^\epsilon u_R\; [1+{\cal O}(u_R)]
\end{equation}
between the $\Phi^4$ coupling constant $u$ and its renormalized
counterpart $u_R$. Here $\mu$ is the inverse length scale that
determines the renormalization flow, and the ${\cal
O}(u_R)$-contributions contain pole terms in $\epsilon$.
The stress
tensor (\ref{tkl}) is already renormalized
\cite{Brown80}, and quantities such as ${\cal K}$ in Eq.\ (\ref{aeq9})
or the functions
\begin{equation}\label{tau1}
\tau^{(i)}={\cal G}_d^{(i)}\left(r/R_+\, , \rho\,
; \mu r, u_R\right)
\end{equation}
with $i=1,2$ from (\ref{troi}) are finite for $\epsilon \rightarrow 0$
order by order in $u_R$.
Eq.\ (\ref{tau1}) follows from dimensional
analysis, assuming that the boundary condition $a,\,b$ take their
fixed-point form and do not introduce any new lengths. If, in addition,
the coupling $u_R$ equals its critical fixed-point value
\begin{equation}\label{usta}
u_R^{*} = \epsilon\:\frac{N+2}{N+8} + {\cal O}(\epsilon^2)\>,
\end{equation}
the renormalization-group equation \cite{Brown80} implies that
the $\tau^{(i)}$ are independent of $\mu r$. By
Eqs.\ (\ref{troi})--(\ref{trel}) and the vanishing trace, the
$\tau^{(i)}$ are also independent of $r/R_+$ and $i$.
Thus, we find that
\begin{equation}\label{taui}
{\cal G}_d^{(i)} \left(r/R_+\, , \rho\, ; \mu r,
u_R^{*}\right) = {\cal G}^*_{d}(\rho)
\end{equation}
only depends on $\rho$, $d$ and the
surface universality classes $a,\,b$.
This provides another derivation of (\ref{tenn}).
\setcounter{equation}{0}
\section{Situations with broken symmetry: Mean field
approximation}\label{3}
In this section
we consider the surface universality classes $(a,\,b) =
(\uparrow, \uparrow),\, (\uparrow, \downarrow), \,(\uparrow,\! O)$ and
$(\uparrow, \! SB)$, where the order parameter symmetry is {\it broken}
at the bulk critical point. With applications to the liquid-gas or
consolute critical points in mind, we only consider a one-component
order parameter. In the standard approach, the order parameter
\begin{equation}\label{spli}
\Phi ({\rm \bf r}) = \langle \Phi ({\rm \bf r})\rangle +
\varphi ({\rm \bf r})
\end{equation}
is decomposed
into a non-vanishing average $\langle \Phi \rangle$
and the fluctuation $\varphi$ around it. Then a
fluctuation (or loop) expansion is carried out, which amounts to
an expansion in powers of $u$.

To leading order one may neglect fluctuations altogether and determine
$\left\langle \Phi \right\rangle\equiv \left\langle \Phi
\right\rangle^{(0)}$ by minimizing $\cal H$ and ignore $\varphi$. In
the concentric geometry $\left\langle \Phi \right\rangle$ only
depends on the distance $r$ from the center and
\cite{G+R}
\begin{equation}\label{mfeq}
 \ddot{m}+\frac{d-1}{r}\dot{m}=m^3\>,
\end{equation}
with
\begin{equation}
m=\sqrt{u/6} \>\>\langle\Phi\rangle^{(0)}
\end{equation}
which has the dimension of an inverse length. For the trace of the
stress tensor one finds
\begin{equation}\label{trst}
\sum_k\langle T_{kk}({\rm \bf r}\,)\rangle_{{\rm \scriptscriptstyle
CON}}^{(0)}=-\frac32\;\epsilon\; m^4/u\>,
\end{equation}
while the radial component is given by
\begin{equation}\label{trad}
r^d\,\langle T_{nn}({\rm \bf r})\rangle_{{\rm \scriptscriptstyle
CON}}^{(0)}= \frac32\,I_d/(u r^{\epsilon})\>,
\end{equation}
where
\begin{equation}\label{fiin}
I_d=\left(2\dot m^2-m^4\right)\,r^4+2\,(d-2)\,m\,\dot m\,r^3\>.
\end{equation}
Here $\dot m\equiv {\rm d}m/{\rm d}r$, and the
superscript zero denotes the $1/u$ contribution to $\left\langle
T\right\rangle$, which is obtained by neglecting fluctuations. The
first and second term on the
right hand side of (\ref{fiin}) come from the
canonical tensor and the improvement term, respectively.

The quantities (\ref{trst}) and (\ref{trad}) determine the $1/u$
contributions to the
right and left hand side of Eq.\,(\ref{trel}), respectively,
and allow one to explicitly check (\ref{trel})
to this order. Note that for
$\epsilon \rightarrow 0$ with $u>0$ fixed the trace vanishes. Thus
$I_4$ is independent of $r$ and represents a first integral of the
differential equation (\ref{mfeq}) if $d=4$.

This integral was encountered in Ref. \cite{G+R}, where it
is pointed out that Eq.\,(\ref{mfeq}) is a special case of the
generalized Emden--Fowler (GEF) equation \cite{Leach}.
For $d=4$, (\ref{mfeq}) belongs to a subclass of the GEF equation, for
which the first integrals are known explicitly \cite{SaBa,Leach}.

A first integral for the mean-field
equation in the spherically symmetric geometry
may also be constructed by means of the stress
tensor if $u \Phi^4$ in (2.20) is replaced by an arbitrary
monomial interaction $u \Phi^M$. The leading contribution in $u$ of
the stress tensor always has a vanishing trace at the upper critical
dimension $d=2M/(M-2)$ (see Ref. \cite{Brown80}). All the
corresponding mean-field equations belong to the above-mentioned
subclass.

Now we assume that the boundary conditions for $m$ at $r=R_-$ and
$r=R_+$ take their fixed-point forms. Then the dimensionless quantity
$I_d$ in (\ref{fiin}) can only depend upon two ratios, say
\begin{equation}\label{twor}
I_d=I_d(r/R_+,\,\rho)
\end{equation}
of the three lengths $r,\,R_-,\,R_+$. In particular, for $d=4$
\begin{equation}\label{orat}
I_4=I(\rho)\>,
\end{equation}
i.e., $I_4$ depends only on $\rho$. Note that the
term in $r^d\,\langle T_{nn}
\rangle_{\scriptscriptstyle{\rm CON}}$ of leading order in $u_R$,
which follows from
substituting (\ref{twor}) and (\ref{relu}) into (\ref{trad}), does
indeed have the form (\ref{tau1}).

The scaling function $Y(\rho)$ in (\ref{tenn}) follows
from $\langle T_{nn}\rangle_{\scriptscriptstyle{\rm CON}}$ for
$u=u_R^*$. Substituting (\ref{relu}), (\ref{usta}) and (\ref{orat})
into (\ref{trad}) and keeping only the leading
$\epsilon$-dependence, one finds
\begin{equation}\label{yrho}
Y(\rho) = \frac{1}{\epsilon}\; \frac{9}{2\pi^2}\;I(\rho)+{\cal
O}(\epsilon^0)\>.
\end{equation}

To obtain solutions $m$ in $d=4$ with the first integral $I_4 \equiv I$,
we follow the procedure in Ref. \cite{G+R}. With the
substitution
$m=v/r$ Eq.\,(\ref{fiin}) becomes
\begin{mathletters}\label{dif}
\begin{equation}\label{dife}
{\rm d}\,{\rm ln}\, r = \pm \sqrt{2} \>\>{\rm d}v\>/\>W(v)\>,
\end{equation}
with
\begin{equation}\label{dife2}
W(v)=\sqrt{I+2v^2+v^4}\>,
\end{equation}
\end{mathletters}\noindent
and solutions can be obtained in terms of elliptic integrals.\\
\underline{Case (i):} For $(a,b)=(\uparrow ,\uparrow)$ the order
parameter profile $m$ tends to $+\infty$ near the two surfaces and has
one minimum in between. This implies $I<0$ and a solution with the
parametric form \cite{G+R}
\begin{mathletters}\label{sol1}
\begin{equation}
v=\frac{\sqrt{\gamma-1}}{\cos \varphi}
\end{equation}
\begin{equation}\label{pr++}
{\rm ln}\>\frac{r}{R_M} = \gamma^{-1/2}\;
F\left(\varphi, \left[(1+1/\gamma)/2 \right]^{1/2}\right)\>,
\end{equation}
\end{mathletters}\noindent
where $F$ is the elliptic integral
\begin{equation}\label{elli}
F(\varphi,k)=\int_0^{\varphi}{\rm
d}\widehat\varphi\,\left(1-k^2\sin^2\!{\widehat\varphi}\right)^{-1/2}\>,
\end{equation}
and
\begin{equation}\label{gamm}
\gamma=\sqrt{1-I}\>.
\end{equation}
Here $r=R_M$
is the radius with minimum $v$, corresponding to $\varphi=0$.
At $\varphi=\pm \pi/2$ where the profile diverges, $r$ takes the
value $r=R_{\pm}$. Eq.\,(\ref{pr++})
implies $R_-R_+=R_M^2$ and
\begin{mathletters}\label{f1}
\begin{equation}\label{f1ln}
{\rm ln}\,\rho=-f_1(I)
\end{equation}
with
\begin{equation}\label{f1de}
f_1(I)=2\gamma^{-1/2}\,K\left(\left[(1+1/\gamma)/2\right]^{1/2}\right)
\>.
\end{equation}
\end{mathletters}\noindent
Here $\rho$ is from (\ref{rhof}), and $K$ is the complete elliptic
integral
\begin{equation}\label{coel}
K(k)=F(\pi/2,k)\>.
\end{equation}
Eqs.\,(\ref{f1}), (\ref{gamm}), and (\ref{yrho}) determine the
scaling function
$Y(\rho)$ for the stress tensor in (\ref{tenn}). Note that $I$
varies from $0$
to $-\infty$ as $\rho$ varies from $0$ to $1$.\vspace{1mm}\\
\underline{Case (ii):} The solution for $(a,b)=(SB,\uparrow)$ or
$(\uparrow ,SB)$ is readily obtained from the solution
in (i). Identifying $R_M$ in (\ref{pr++}) with
the radius
of the $SB$ boundary, i.e. considering
\begin{mathletters}\label{rre}
\begin{equation}\label{rre1}
R_M=R_-<r<R_+
\end{equation}
or
\begin{equation}\label{rre2}
R_-<r<R_M=R_+
\end{equation}
\end{mathletters}\noindent
leads to the relation
\begin{equation}\label{f1sb}
{\rm ln}\rho=-\frac12\>f_1(I)\>,
\end{equation}
which only differs from (\ref{f1ln}) by a factor $1/2$. The reason for
the identifications
(\ref{rre}) can be understood by conformally transforming the
mean-field
(MF) solution (\ref{sol1}) for $(\uparrow ,\uparrow)$ in concentric
geometry to spheres of equal size (SES)
in ${\rm \bf r}^{\,\prime}$-space, with the mid-plane
$r^{\,\prime}_{\bot}=0$ corresponding to the sphere with $r=R_M$
(see the remark just below Eq.\,(\ref{inte})).
The transformed MF solution
is invariant under $r_{\bot}^{\,\prime}\rightarrow
-r_{\bot}^{\,\prime}$, and the normal derivative vanishes at
$r_{\bot}^{\,\prime}=0$. Since the MF equation
$\Delta\langle\Phi\rangle^{(0)}=
(u/6)\left(\langle\Phi\rangle^{(0)}\right)^3$ is conformally invariant
in $d=4$, this is the correct MF solution for a half--space
$r_{\bot}^{\,\prime}>0$
in $d=4$ with a planar $SB$ surface at $r_{\bot}^{\,\prime}=0$ and a
spherical hole
with $\uparrow$ boundary conditions. Transforming this solution back
to the concentric geometry leads to (\ref{rre}) and (\ref{f1sb}). Note
that for the two geometries (\ref{rre}) the radial derivative
$\mbox{d}\,{\rm ln}\, m(r)/\mbox{d}r$
at the $SB$ surfaces does not vanish but equals
$-1/R_-$ and $-1/R_+$ for (\ref{rre1}) and (\ref{rre2}), respectively
\cite{foo8a}.
Thus the MF order-parameter profile increases (decreases) on
approaching the convex (concave) $SB$
surface from the interior of the critical system,
which is reasonable. The MF values $\pm 1$
for the ratio of the logarithmic normal derivative and the inverse
radius of curvature in $d=4$ are consistent with results in Ref.
\cite{McAOsII}.
\vspace{2mm}\\
\underline{Case (iii):} For $(a,b)=(\downarrow ,\uparrow)$,
the profile tends to
$\pm\infty$ for $r\rightarrow R_{\pm}$ and has a zero in between. This
implies
$I>0$. Though the $I$-dependence of $m(r,I)$ is perfectly regular at
$I=1$, it is advantageous to choose different
parametric representations for the intervals
$0<I<1$ and $I>1$, respectively. They are given by
\begin{mathletters}\label{sol2}
\begin{equation}\label{sol21}
v=(1-\gamma)^{1/2}\,\tan\varphi\>,
\end{equation}
with
\begin{equation}\label{sol22}
{\rm
ln}\,\frac{r}{R_0}=\left(\frac{2}{1+\gamma}\right)^{1/2}\,F\left(
\varphi,
\left[2/(1+1/\gamma)\right]^{1/2}\right)
\end{equation}
\end{mathletters}\noindent
for $0<I<1$ (with $\gamma$ from (\ref{gamm})) and by
\begin{mathletters}\label{sol3}
\begin{equation}\label{sol31}
v=I^{1/4}\,\tan\varphi\>,
\end{equation}
with
\begin{equation}\label{sol32}
{\rm ln}\;\frac{r}{R_0}=2^{-1/2}\;I^{-1/4}\;F\left(2\varphi,
\left[(1-I^{- 1/2})/2\right]^{1/2}\right)
\end{equation}
\end{mathletters}\noindent
\noindent
for $I>1$. The profile diverges at $\varphi=\pm\pi/2$, corresponding
to
$r=R_{\pm}$, and has a zero at $\varphi=0$, corresponding to
$r=R_0=\left(R_+R_-\right)^{1/2}$.
Eqs.\,(\ref{sol22}) and (\ref{sol32}) determine the relation
$\rho=\rho(I)$.
A form which holds for both $0<I<1$ and $I>1$ is
\begin{mathletters}\label{f3}
\begin{equation}\label{f3en}
{\rm ln}\rho=-f_3(I)\>,
\end{equation}
with
\begin{equation}\label{f3de}
f_3(I)=2^{3/2}\>I^{-1/4}\;\frac{2}{\delta}\>K\left(\frac{2}{\delta}-
1\right)\>.
\end{equation}
\end{mathletters}\noindent
Here
\begin{equation}\label{delt}
\delta=1+\left[\left(1+I^{-1/2}\right)/2\right]^{1/2}\>.
\end{equation}
\vspace{1mm}\\
\underline{Case (iv):}
The solution for $(a,b)=(O,\uparrow)$ follows from that
in case (iii) by identifying $R_0$ with the radius $R_-$ of the
$O$ boundary, where the order parameter vanishes \cite{Diehl86}.
This implies
\begin{equation}\label{f3do}
{\rm ln}\rho =-\frac12\,f_3(I)\>.
\end{equation}

The four functions $\rho=\rho(I)$ in Eqs.\,(\ref{f1}), (\ref{f1sb}),
(\ref{f3}), and (\ref{f3do}) determine via (\ref{yrho}) the scaling
function
$Y(\rho)$ of the stress tensor and via (\ref{delf}), (\ref{fcon}),
and
(\ref{deef}) the free energy and the Casimir forces for the four pairs
$(\uparrow ,\uparrow)$, $(SB,\uparrow)$, $(\downarrow ,\uparrow)$,
$(O,\uparrow)$ of boundary conditions.
Integrating according to (\ref{deef}) yields the parametric form
\begin{equation}\label{fkin}
{\cal F}\left(\kappa(\rho(I))\right)
=\frac{9\;S_d}{2^{d+1}\,\pi^2\,\epsilon}\>
\int_0^{I}\>{\rm d} \widehat I\>\>\frac{\mbox{d}\,
\mbox{ln}\,\rho(\widehat I)}{\mbox{d}\widehat I}\:\widehat I
\end{equation}
for the scaling function,
with $\kappa=\kappa(\rho)$ from (\ref{kap3}) and $\rho=\rho(I)$
defined as above. On varying $\rho$ or $\kappa$, ${\cal F}$,
$\>\langle T_{nn}\rangle_{\scriptscriptstyle \rm CON}$, and $I$
remain negative in the cases
(i) and (ii) and positive in the cases
(iii) and (iv), implying attractive and repulsive
interactions, respectively.
The remaining integration in (\ref{fkin}) has been carried out
numerically. Results for the scaling function for the four
combinations of boundary conditions are shown in Fig.\ 2.
Since the prefactor in (\ref{fkin}) behaves like
$1/\epsilon$ as $d\rightarrow 4$, we have plotted
the finite quantity
$\tilde {\cal F}= \lim_{d\rightarrow 4}\: (4-d)\, {\cal F}$.

It is instructive to check that our solutions exhibit the
limiting
behavior discussed in Subsecs.\ \ref{2B} and
\ref{2C}. Consider first
the limit $\rho\rightarrow 1$,
where $|I|\rightarrow \infty$. In this case
\begin{equation}\label{li13}
(f_1,\,f_3)\rightarrow
\left(2|I|^{-1/4}\;,\>2^{3/2}I^{-1/4}\right)\cdot
K(1/\sqrt{2})\>.
\end{equation}
Substituting the corresponding $I(\rho)$ into (\ref{yrho}), one
finds that $Y(\rho)$ does indeed have the behavior (\ref{limy})
with $d=4$ and that
\begin{mathletters}\label{de}
\begin{equation}\label{depp}
\Delta_{\uparrow\uparrow}=-\frac{1}{\epsilon}\;\frac{3}{2\pi^2}
\left[K\left(1/\sqrt{2}\right)\right]^4
\end{equation}
and
\begin{equation}\label{deot}
\Delta_{SB\,\uparrow}=\Delta_{\uparrow\uparrow}/16\;,\quad
\Delta_{\downarrow\uparrow}=-4\,\Delta_{\uparrow\uparrow}\;,\quad
\Delta_{O\uparrow}=\Delta_{\downarrow\uparrow}/16\>.
\end{equation}
\end{mathletters}\noindent
The expressions in (\ref{de}) reproduce the known
results \cite{InNi86,EiSt} for the Casimir amplitudes
in the parallel-plate geometry for the symmetry-breaking
cases and $d$ near $4$.

Next consider $\rho\rightarrow 0$, where $|I|\rightarrow 0$.
In this case
\begin{mathletters}\label{li}
\begin{equation}\label{limo}
\left(f_1,f_3\right)\rightarrow -{\rm ln}\left(|I|/64\right)
\end{equation}
which implies
\begin{equation}\label{limi}
I\rightarrow 64\cdot\left[ -\rho,\,-\rho^2,\,\rho,\,\rho^2\right]
\end{equation}
\end{mathletters}\noindent
in cases (i)-(iv). In order to compare with (\ref{ylim})
we need the expressions
\begin{equation}\label{dra4}
\frac{A_a^{\Psi}\,A_b^{\Psi}}{B_{\Psi}}=\frac{36}{\epsilon}\;
\left[\;1\;,\>\frac12\;,\>-1\;,\>-\frac12\;\right]\>\>,
\quad d\rightarrow 4
\end{equation}
for $AA/B$ in our four cases. Here $\Psi=[\Phi,\Phi^2,\Phi,\Phi^2]$
as
noted just below Eq.\,(\ref{frco}). The results (\ref{dra4}) follow
from the behavior
\begin{equation}\label{leau}
\pm\sqrt{\frac{12}{uz^2}}\>\>,\qquad\frac{12}{uz^2}\>,
\end{equation}
to leading order in $u$
of $\langle\Phi\rangle$, $\langle\Phi^2\rangle$ in the
half--space with
$a=\uparrow$ or $\downarrow$
(see Eq.\ (\ref{ampl})), from the corresponding behavior
\begin{equation}\label{cobe}
\mp \tilde S_d (2z)^{2-d}
\end{equation}
of $\langle\Phi^2\rangle$ in the half--space
with an $a=O$ or $SB$ surface,
and from the leading-order contributions
\begin{equation}\label{dokn}
\tilde S_d\;r^{2-d}\>\>,\qquad2\left(\tilde S_d\;r^{2-d}\right)^2
\end{equation}
to the bulk correlation functions $\langle\Phi\,\Phi\rangle$,
$\langle\Phi^2\,\Phi^2\rangle$ (cf. Eq.\ (\ref{corr})).
Here
\begin{equation}\label{tisd}
\tilde S_d=\frac14\,\pi^{-d/2}\Gamma\left(\frac{d}{2}-1\right)
=\frac{1}{(d-2)\,S_d}\>,
\end{equation}
and $u$ must be
replaced by (\ref{relu}) with the fixed-point value (\ref{usta})
with $N=1$
for $u_R$. Using the results (\ref{limi}) and (\ref{dra4}),
one finds by
means of Eq.\,(\ref{yrho}) and from $x_{\Phi}\rightarrow 1$,
$x_{\Phi^2}\rightarrow 2$ for $d\rightarrow 4$ that our four solutions
are indeed consistent with (\ref{ylim}).

Now we turn to the
density profiles and their behavior near a boundary. Consider
first the $r\rightarrow R_+$ behavior of $m$ in the concentric
geometry for $d=4$ with universality class $b=\uparrow$
for the boundary at $R_+$.
This follows directly from the differential
equation (\ref{dife}), which implies
\begin{equation}\label{lnrr}
{\rm ln}\>\frac{R_+}{r}=\sqrt{2}\int_v^{\infty}
{\rm d}\widehat v\,/\,W(\widehat v)\>.
\end{equation}
Here we have used $v(R_+)=\infty$ and assumed that $v$
is larger
than a possible minimum in $v(r)$. For $r$ close to $R_+$, $v$
tends to infinity, and one may expand the integrand in (\ref{lnrr}) in
powers of $1/\widehat v$. The integration is elementary, and one finds
\begin{equation}\label{fidg}
m_{a,\uparrow}(r)=m_{\uparrow}^{({\rm fc})}(r)\left[
1-\frac{1}{40}\;I_{a,\uparrow}\>\left(\frac{R_+-r}{R_+}\right)^4+
\ldots\right]\>.
\end{equation}
The first factor on the right hand
side is the profile inside a full spherical
container with radius $R_+$, which follows from (\ref{lnrr}) with $I=0$.
The second term in square brackets respresents the leading correction
for $r\rightarrow R_+$ due to the inner spherical
boundary $a$ with radius $R_-$. The prefactor $I_{a,\uparrow}$
denotes $I$ in the above cases (i), (ii), (iii), or (iv),
with $a=\uparrow$, $SB$, $\downarrow$,
or $O$ (compare Eq.\ (\ref{limi})).
The result (\ref{fidg}) also follows from the
parametric representations (\ref{pr++}), (\ref{sol22}), (\ref{sol32})
for $m(r)$.

The corresponding expansion near the planar wall with universality
class $b=\uparrow$ in the SPW geometry introduced below (\ref{inte})
follows from substituting (\ref{fidg}) into the transformation
formula
\begin{equation}\label{trfo}
m_{{\rm \scriptscriptstyle S}_a
{\rm \scriptscriptstyle PW}_{\uparrow}}({\rm \bf
r}^{\,\prime})=\left[b({\rm \bf r}^{\,\prime})
\right]^{-1}\>m_{a,\uparrow}(r)
\end{equation}
for mean-field solutions in $d=4$. Here $b({\rm \bf r}^{\,\prime})$
is from (\ref{losc}) with $R=R_+$. This leads to the result
\begin{equation}\label{expw}
m_{{\rm \scriptscriptstyle S}_a
{\rm \scriptscriptstyle PW}_{\uparrow}}({\rm \bf r}^{\,\prime})/
m_{{\rm \scriptscriptstyle PW}_{\uparrow}}
(r_{\bot}^{\,\prime})= 1-
\frac{1}{40}\,I_{a,\uparrow}\;{\scriptstyle\cal Z}^4
+\ldots\>\>,
\end{equation}
with
\begin{equation}\label{calz}
{\scriptstyle\cal Z}=4\, r_{\bot}^{\,\prime}\,R_+/\left[({\rm \bf
r}_{\parallel}^{\,\prime})^2
+(2R_+)^2\right]\>,
\end{equation}
for the leading behavior of $\left[R_+-r({\rm \bf
r}^{\,\prime})\right]/R_+$ as
$r_{\bot}^{\,\prime}\rightarrow 0$. Here we use the
relation between the profiles
$m_{PW_{\uparrow}}$ in half--space and $m_{\uparrow}^{\rm (fc)}$ inside
a full
spherical container that follows from the conformal transformation
(\ref{trfo}) \cite{B+E}. Using Eqs.\,(\ref{tspw}), (\ref{rhod})
and (\ref{yrho}) for
$\langle T_{\!\bot\!\bot}\rangle_{{\rm \scriptscriptstyle SPW}}$ and
the result \cite{EiSt}
\begin{equation}\label{sham}
C_{\Phi}^{(\uparrow)}=\epsilon\,\frac{4\pi^2}{45}\>+{\cal
O}(\epsilon^2)
\end{equation}
for the universal short-distance amplitude, we have checked that
Eqs.\ (\ref{expw}), (\ref{calz}) are fully consistent with the
short-distance expansion
(\ref{shor}) when applied to a
$\langle \Psi\rangle_{{\rm \scriptscriptstyle SPW}}=
\langle\Phi\rangle_{{\rm \scriptscriptstyle SPW}}$
average for $d\rightarrow 4$.

In the other cases such as $(\Psi,\,b)=(\Phi^2,\,O)$ with $a=\uparrow$,
the short
distance expansion (\ref{shor})
about the planar wall with universality class $b$ can be
checked in a similar way \cite{foo8b}. The short-distance coefficient
$C_{\Phi^2}^{(O)}$
turns out \cite{EKD} to be of order $\epsilon^0$ (see Eq. (\ref{cphi})
below).
As a consequence, the first term
in square brackets in (\ref{shor}) leads to a
contribution which is smaller by
a factor $\epsilon$ than the contribution from the second term
and is not present in the mean-field solution.

In Appendix \ref{B1} we evaluate density profiles
for $R_-\rightarrow 0$ for arbitrary fixed $0<r<R_+$.

For $\rho\rightarrow 1$, the density profile in the
parallel-plate geometry \cite{InNi86} is obtained.
This has been shown for
$(a,\,b)=(\uparrow,\uparrow)$ in Ref.\,\cite{G+R}, and similar
arguments apply for other $a,\>b$.
\setcounter{equation}{0}
\section{Situations with symmetry preserved: Gaussian model}\label{4}
The boundary conditions
$(a,\,b)=(O,\,O),\;(O,\,SB),\;(SB,\,SB)$
do not destroy \cite{Diehl86} the bulk $O(N)$ symmetry of the
$N$-component
order parameter at the bulk critical point.
In these cases the starting point of the $d=4-\epsilon$ expansion
is \cite{Amit} the critical Gaussian model. This is defined by
Eq.\,(\ref{hbul})
with $u=0$. It is instructive \cite{foo9} to consider the critical
Gaussian
model for arbitrary $d>2$. For all $d$ it is at its critical fixed
point, i.e. dilatation and conformally invariant on all length scales.
This also holds in the presence of boundaries $a$ and $b$ if the
boundary conditions take their fixed-point forms.

In the concentric geometry the Gaussian order-parameter correlation
functions
or `propagators' with $a,b\in\left\{O,SB\right\}$,

\begin{equation}\label{prop}
\langle \Phi_i({\rm
\bf r}_1)\Phi_j({\rm \bf r}_2)\rangle_{\rm \scriptscriptstyle
CON}=\delta_{ij}\:G_{ab}({\rm \bf r}_1,{\rm \bf r}_2)=
\delta_{ij}\:\tilde G_{ab}(r_1,r_2;\eta)\>,
\end{equation}
only depend on the positions ${\rm \bf r}_1,{\rm \bf r}_2$
through their distances $r_1,r_2$
from the center and the cosine of the enclosed angle
\setcounter{equation}{0}
\begin{mathletters}\label{pro}
\begin{equation}\label{etad}
\eta={\rm \bf r}_1\cdot{\rm \bf r}_2/(r_1r_2)\>.
\end{equation}
\end{mathletters}\noindent
Here $i$ and $j$ denote components of the order parameter.
One finds
\begin{equation}\label{prex}
\tilde G_{ab}(r_1,r_2;\eta)=\tilde S_d\;\left(r_<\:
r_>\right)^{-\vartheta}
\>\sum_{l=0}^{\infty}
C_l^{(\vartheta)}(\eta)\>\>W^{(l)}_{a,b}\>\;
{\chi}_a\!\left((r_</R_-)^{\lambda}\right)
{\chi}_b\!\left((r_>/R_+)^{\lambda}\right)
\end{equation}
with
\setcounter{equation}{1}
\begin{mathletters}\label{def}
\begin{equation}\label{rsrl}
r_<=\min(r_1,r_2)\>,\qquad r_>=\max(r_1,r_2)
\end{equation}
and
\begin{equation}\label{vart}
\vartheta=(d-2)/2\>.
\end{equation}
The $C_l^{(\vartheta)}$ are hyperspherical (Gegenbauer) polynomials
\cite{Abram},
and
\begin{equation}\label{chid}
\chi_O(y)=y-\frac{1}{y}\>,\qquad \chi_{SB}(y)=y+\frac{1}{y}
\end{equation}
and
\begin{equation}\label{lamd}
\lambda=l+\vartheta\>.
\end{equation}
The quantities $W$ depend on the ratio $\rho=R_-/R_+$
of the radii of concentric boundaries and are given by
\begin{eqnarray}\label{wdef}
& W_{O,O}^{(l)}\:&=\:-\left[P^{-1}-P\right]^{-1}\nonumber\\*[2mm]
& W_{O,SB}^{(l)}\:&=\:\left[P^{-1}+P\right]^{-1}\nonumber\\*[2mm]
& W_{SB,O}^{(l)}\:&=\:-\left[P^{-1}+P\right]^{-1}\nonumber\\*[2mm]
& W_{SB,SB}^{(l)}\:&=\:\left[P^{-1}-P\right]^{-1}\>,
\end{eqnarray}
with
\begin{equation}\label{pdef}
P=\rho^{\lambda}\>.
\end{equation}
\end{mathletters}\noindent
The $G_{ab}(\mbox{\bf r}_1,\,\mbox{\bf r}_2)$ remain unchanged for
$\mbox{\bf r}_1$ $\leftrightarrow$ $\mbox{\bf r}_2$
and satisfy $\Delta_{{\rm\bf r}_1}\,G_{ab}=-
\delta(\mbox{\bf r}_1-\mbox{\bf r}_2)$ with
$\Delta$ and $\delta$ the Laplacian and the Dirac $\delta$-function,
respectively, in $d$ dimensions. Note that $\chi_O$ and $\partial
\chi_{SB}/\partial y$ vanish at the boundaries where $y=1$.
This implies the fixed-point boundary conditions
\begin{equation}\label{fibc}
G_{O,b}=0\>,\qquad G_{a,O}=0
\end{equation}
and \cite{McAOsII}
\begin{equation}\label{bcsb}
\partial_{r_<}\,G_{SB,b}=-\frac{\vartheta}{R_-}\,G_{SB,b}\>,\qquad
\partial_{r_>}\,G_{a,SB}=-\frac{\vartheta}{R_+}\,G_{a,SB}\>.
\end{equation}
Here the first and second relations in (\ref{fibc})
and (\ref{bcsb}) hold for $r_<=R_-$ and $r_>=R_+$, respectively.

A conformal transformation
\begin{equation}\label{coga}
G^{({\rm \scriptscriptstyle SPW})}_{ab}\left({\rm \bf
r}_1^{\,\prime},\,{\rm \bf r}_2^{\,\prime}\right)=
\left[b\left({\rm \bf r}_1^{\,\prime}\right)\,
b\left({\rm \bf r}_2^{\,\prime}\right)\right]^{-\vartheta}
G_{ab}\left({\rm \bf r}_1,\,{\rm \bf r}_2\right)
\end{equation}
to the sphere near the planar wall (SPW) geometry
shows that (\ref{bcsb}) is equivalent to the usual Neumann boundary
condition
for a planar $SB$ surface: Consider e.g.\ (\ref{cotr}) with
$R=R_+$, which transforms the outer concentric boundary
$b=SB$ to the plane $r^{\,\prime}_{\bot}=0$ (see Fig.\ 1).
To prove that the derivative vanishes
at $r_{1\bot}^{\,\prime}=0$ for
$r_{2\bot}^{\,\prime}>0$, we show that
the continuation in $r_{1\bot}^{\,\prime}$
of $G^{({\rm \scriptscriptstyle SPW})}$ across the
boundary is even in
$r_{1\bot}^{\,\prime}$. The points corresponding to
${\rm \bf r}_{1}^{\,\prime}=({\rm \bf r}_{1\parallel}^{\,\prime}\:,\:
{r}_{1\bot}^{\,\prime}$) and its mirror image
${\rm \bf r}_{1}^{\,\prime\,(I)}=({\rm \bf
r}_{1\parallel}^{\,\prime}\:,
-{r}_{1\bot}^{\,\prime})$
are ${\rm \bf r}_1$ and
\begin{equation}\label{rima}
{\rm \bf r}_{1}^{\>(I)}={\rm \bf r}_{1}\>R_+^2/r_1^2\>,
\end{equation}
respectively,
as follows from inversion about the sphere $r=R_+$. Since
\begin{equation}\label{chsb}
\chi_{SB}\!\left((r_1^{(I)}/R_+)^{\lambda}\right)=
\chi_{SB}\!\left((r_1/R_+)^{\lambda}\right)
\end{equation}
and
\begin{equation}\label{rim2}
r_1^{(I)}\,b\!\left({\rm \bf r}_1^{\>\prime\,(I)}\right)
=r_1\,b\!\left({\rm \bf r}_1^{\,\prime}\right)\>,
\end{equation}
Eq.\ (\ref{coga}) shows that
$G^{({\rm \scriptscriptstyle SPW})}$ is
indeed even in $r_{1\bot}^{\,\prime}$.

An inversion about the sphere $r=\sqrt{R_- R_+}$ exchanges the
boundaries
$a$ and $b$ and implies
\begin{equation}\label{geba}
\tilde G_{ba}(r_1,r_2;\eta)=
\left(\frac{R_- R_+}{r_1 r_2}\right)^{2\vartheta}\>\,
\tilde
G_{ab}\!\left(\frac{R_-R_+}{r_1},\frac{R_-R_+}{r_2};\eta\right)\>.
\end{equation}
One easily verifies that $\tilde G_{O,SB}$, $\tilde G_{SB,O}$ in
(\ref{prex}) are consistent with (\ref{geba}).

$G_{ab}$ in (\ref{prex}) should be compared with the propagator
in the unbounded bulk
\begin{equation}\label{gbul}
G_{\rm bulk}({\rm \bf r}_1,\,{\rm \bf r}_2)=\tilde S_d\>|{\rm \bf
r}_1-{\rm \bf r}_2|^{-
2 \vartheta}=\tilde S_d\>
(r_<\:r_>)^{-\vartheta}\,\sum_{l=0}^{\infty}
C_l^{(\vartheta)}(\eta)\,(r_</r_>)^{\lambda}\>.
\end{equation}
This equation follows from $G_{ab}$ in (\ref{prex}) in the limits
$R_-\rightarrow 0$, $R_+\rightarrow\infty$ with ${\rm \bf r}_1,\,{\rm
\bf r}_2$ fixed.
Below we will need the difference
\begin{eqnarray}\label{dege}
G_{ab}({\rm \bf r}_1,{\rm \bf r}_2)-
G_{\rm bulk}({\rm \bf r}_1,{\rm \bf r}_2)&=&
\tilde S_d\>\sum_{l=0}^{\infty}\,C_l^{(\vartheta)}(\eta)
\left[\sigma_{ab}^{(1)}P^{-2}-1\right]^{-1}\cdot\nonumber\\
& &\left\{\frac{r_1^l}{r_2^{\vartheta+\lambda}}+
\frac{r_2^l}{r_1^{\vartheta+\lambda}}+
\sigma_{ab}^{(2)}\:\frac{(r_1r_2)^l}{R_-^{2\lambda}}
+\sigma_{ab}^{(3)}\frac{R_+^{2\lambda}}{(r_1r_2)^{\vartheta+\lambda}}
\right\}\>,
\end{eqnarray}
which is regular for ${\rm \bf r}_1\rightarrow {\rm \bf r}_2$ at
interior points.
Here
\setcounter{equation}{10}
\begin{mathletters}
\begin{eqnarray}\label{sigd}
\sigma_{ab}^{(1)}&=&(1,-1,-1,1)\nonumber\\*[2mm]
\sigma_{ab}^{(2)}&=&(-1,-1,1,1)\nonumber\\*[2mm]
\sigma_{ab}^{(3)}&=&(-1,1,-1,1)
\end{eqnarray}
for
\begin{equation}\label{abde}
a\,b=(O\:O,\:O\:SB,\:SB\:O,\:SB\:SB)\>.
\end{equation}
\end{mathletters}
\noindent

In order to
calculate $\langle T_{nn}(r)\rangle_{\rm \scriptscriptstyle CON}$ in
the
Gaussian model, we use (\ref{tekl}) with $u=0$
in the Hamiltonian (\ref{hbul}). Since the average
$\langle T_{kl}\rangle_{\rm bulk}$ in the unbounded bulk
vanishes, we evaluate the averages $\langle \Phi \Phi\rangle_{\rm
\scriptscriptstyle CON}$
and $\langle \Phi^2\rangle_{\rm\scriptscriptstyle CON}$ that arise
from Eq.\,(\ref{tekl})
by using $G_{ab}-G_{\rm bulk}$ from (\ref{dege})
instead of $G_{ab}$.
This yields
\begin{eqnarray}\label{tgau}
N^{-1}\,\lefteqn{\langle T_{nn}({\rm \bf r})
\rangle_{\rm \scriptscriptstyle CON}= }\nonumber\\
& & \hspace{5mm}-  r^{-d}\;\tilde S_d\>
\sum_{l=0}^{\infty}\left[\sigma_{ab}^{(1)}\,P^{-2}-1\right]^{-1}
{d-3+l \choose l}
\left\{2l\,(d-2+l)+\frac12 (d-2)^2\right\}\>.
\end{eqnarray}
Using (\ref{deef}), one eventually obtains
\begin{equation}\label{nmof}
N^{-1}\:{\cal F}(\kappa)=\frac{1}{d-2}
\sum_{l=0}^{\infty}\>{\rm
ln}\left(1-\sigma_{ab}^{(1)}\,\rho(\kappa)^{2\lambda}
\right)\:\frac{1}{2\,\lambda}
\:
{d-3+l \choose l}
\left\{2l\,(d-2+l)+\frac12 (d-2)^2\right\}\>
\end{equation}
where the relation between $\kappa$ and $\rho$ is given
in (\ref{kap3}).
The signs of both ${\cal F}$ and $\langle T_{nn}
\rangle_{\scriptscriptstyle\rm CON}$ are opposite to
the sign of $\sigma^{(1)}_{ab}$.
This holds for all values of $\kappa$ or $\rho$. Thus
the interaction is always attractive if $a=b$
and repulsive if $a\neq b$. We have evaluated numerically
the sum in (\ref{nmof}). The function
$-{\cal F}_{O,O}=-{\cal F}_{SB.SB}$ for $d=4$ is shown
in Fig.\ 2.

The profile of the energy density \cite{foo3b} is obtained from
$G_{ab}-G_{\rm bulk}$. It takes the form
\begin{eqnarray}\label{endi}
\lefteqn{N^{-1}\,\langle \roarrow  \Phi^{\>2}({\rm \bf r})
\rangle_{\rm \scriptscriptstyle CON}= }\nonumber\\
& & \hspace{5mm}  r^{2-d}\,\tilde S_d
\sum_{l=0}^{\infty}\left[\sigma_{ab}^{(1)}\,P^{-2}-1\right]^{-1}
{ d-3+l \choose l}
\left\{2+\sigma_{ab}^{(2)}\;\left(\frac{r}{R_-}\right)^{2\lambda}
+\sigma_{ab}^{(3)}\;\left(\frac{R_+}{r}\right)^{2\lambda}\right\}\>.
\end{eqnarray}

Note the different sources of $r$-dependence in the expression
(\ref{endi}) for the energy density. The $r$-dependence is
very similar in each of the contributions
to $\langle T_{nn}\rangle$ coming from Eqs.\,(\ref{tkl}).
However,
$\left(r/R_-\right)^{2\lambda}$ and $\left(R_+/r\right)^{2\lambda}$
cancel in the sum, and
$\langle T_{nn}\rangle$ in (\ref{tgau}) turns out proportional
to $r^{-d}$, in agreement with (\ref{tenn}).
The $(\partial\Phi)^2$ terms with $\partial$
parallel or perpendicular to ${\rm \bf r}$
that arise from the canonical tensor in
(\ref{tekl}) contribute half of the first term in the curly brackets
on the right hand side of (\ref{tgau}), while the second
term comes from the improvement
${\cal J}$ in (\ref{impr}).  The sum of the two terms in curly
brackets equals $2\lambda^2$.

{}From (\ref{tgau}) and (\ref{endi}), one may again check that the two
combinations $(a,b)=(O,\,SB)$ and $(SB, \, O)$ are related by an
inversion about the sphere $r=\sqrt{R_-\,R_+}$. This operation
preserves $\langle T_{nn}\rangle$, while in
$\langle\roarrow\Phi^{\>2}\rangle$ the ratios $r/R_-$ and
$R_+/r$ are exchanged. Note that $\langle T_{nn}\rangle$ is the
same for $(O,\,O)$ and $(SB,\,SB)$, too.
However, the latter identity is limited to the
Gaussian model, as we argue below.

We now consider the limiting behavior of the stress tensor in
(\ref{tgau}). According to the discussion in Subsec.\ \ref{2B},
$\langle T_{nn}\rangle$
should approach the
parallel-plate result when $\rho=1-(L/R_+)$ tends
to $1$. In this case
\begin{equation}\label{plim}
P^{-2}\approx\exp\left(2\lambda L/R_+\right)\>.
\end{equation}
The exponential decay of the first factor in (\ref{tgau}) is
very slow, and the sum in (\ref{tgau}) may be replaced by an
integral
\begin{equation}\label{suli}
\sum_{l}\longrightarrow \int_{0}^{\infty}{\rm d}l
\left[\sigma^{(1)}_{ab}\;P^{-2}-1\right]^{-1}
\;2\:l^{\,d-1}/\Gamma(d-2)\>.
\end{equation}
Here $P^{-2}$ is from (\ref{plim}) with $\lambda$ replaced by $l$, and
the second and third factors in the sum of (\ref{tgau}) have been
replaced by
their limiting forms for $l\gg 1$. The integral in (\ref{suli}) can be
expressed in terms of the Riemann zeta function $\zeta(d)$
\cite{Gradst}.
Comparison with Eqs.\,(\ref{tenn}) and (\ref{limy}) shows that
$\langle T_{nn}\rangle$ does indeed have
the expected limiting form, with the
well-known Casimir amplitudes \cite{Symanzik81,Krech91}
\begin{mathletters}\label{caa}
\begin{equation}\label{caa1}
\Delta_{O,\,O}=\Delta_{SB,\,SB}=-N\left(4\pi\right)^{-d/2}\;
\Gamma\left(\frac{d}{2}\right)\;\zeta(d)
\end{equation}
and
\begin{equation}\label{caa2}
\Delta_{O,\,SB}=-\left(1-2^{1-d}\right)\Delta_{O,\,O}
\end{equation}
\end{mathletters}\noindent
for the $N$-component Gaussian model bounded by plates of type $O$ or
$SB$.

In the other limit $\rho\rightarrow 0$ the leading behavior of
(\ref{tgau}) comes from the $l=0$ term in the sum, which is given by
\begin{equation}\label{rho0}
\sum_{l}\longrightarrow
\sigma_{ab}^{(1)}\;\rho^{2\vartheta}\;\frac12\,(d-2)^2\>.
\end{equation}
This should be compared with the amplitude ratios
\begin{equation}\label{arat}
A_a^{\Phi^2} A_b^{\Phi^2}/B_{\Phi^2}\>=\>\sigma_{ab}^{(1)}\;N/2
\end{equation}
for the combinations (\ref{abde}) of $a$ and $b$. These follow from
Eqs.\,(\ref{cobe}) and (\ref{dokn}). Comparing with (\ref{tenn}) and
(\ref{tisd}), one obtains the expected limiting behavior
(\ref{ylim}) with $x_{\Phi^2}=d-2$.

We conclude the discussion of $\langle T_{nn}\rangle_{
\rm\scriptscriptstyle CON}$ in the limits $\rho\rightarrow 1$ and
$\rho\rightarrow 0$ with two remarks: (i) Only the canonical part of
$T_{kl}$, i.e. the first term in curly brackets in (\ref{tgau}),
contributes in the limit $\rho\rightarrow 1$. This is to be expected,
since the average of the improvement term, which
contains a total derivative parallel to the plates,
vanishes due to translational invariance in the parallel-plate
geometry. It is less obvious that the
leading behavior for $\rho\rightarrow 0$ is only due to the
improvement term (second term in curly brackets in (\ref{tgau})),
as we
have noted in Eq.\,(\ref{rho0}). (ii) For a normal critical system
the $\Phi^4$-interaction of (\ref{hbul}) must be included. Then
$\langle T_{nn}\rangle$ and thus $Y$ and ${\cal F}$ for $(O,\,O)$ and
$(SB,\,SB)$ are no longer the same. From (\ref{ylim})
there should already
be a deviation in first order in $\epsilon=4-d$
for $\rho\rightarrow 0$, since
\begin{equation}\label{e420}
\left({A_{SB}^{\Phi^2}}\;/\;{A_{O}^{\Phi^2}}\right)^2=
1+2\,\epsilon\;\frac{N+2}{N+8}+{\cal O}(\epsilon^2)\>.
\end{equation}
However, for $\rho\rightarrow 1$, i.e. in the
parallel-plate geometry there is no deviation in this order
\cite{Krech91}.

Now we turn to the profile (\ref{endi})
of the energy density. This is the
counterpart of the discussion at the end of Sec.\ \ref{3}. For $r$
fixed and $R_-\rightarrow 0$ only the second term in curly brackets
in Eq.\,(\ref{endi}) survives, and the concentric profile tends to
\begin{eqnarray}\label{enwa}
N^{-1}\,\langle \roarrow  \Phi^{\>2}({\rm \bf r})
\rangle_{\rm \scriptscriptstyle CON}& \longrightarrow &
 r^{2-d}\:\sigma_{ab}^{(3)}\:\tilde S_d\,\sigma_{ab}^{(3)}
\sum_{l=0}^{\infty}
{d-3+l\choose l}
\;\left(\frac{r}{R_+}\right)^{2\lambda}\nonumber\\*[4mm]
& &=\sigma_{ab}^{(3)}\;\tilde S_d\;\left\{R_+\,\left[1-
\left(\frac{r}{R_+}\right)^2\right]\right\}^{-(d-2)}=N^{-1}
\,\langle \roarrow  \Phi^{\>2}({\rm \bf r})\rangle_{\rm fc}\>,
\end{eqnarray}
which is the profile inside a full spherical container
(fc) with radius $R_+$ and universality class $b$
(see the corresponding general expression in Ref.\,\cite{B+E}).
Below we will need the difference
\begin{eqnarray}\label{eddi}
\lefteqn{N^{-1}\,\left[\langle \roarrow  \Phi^{\>2}({\rm \bf r})
\rangle_{\rm \scriptscriptstyle CON} -
\langle \roarrow  \Phi^{\>2}({\rm \bf r})\rangle_{\rm fc} \right]=
}\nonumber\\*[3mm]
& & \hspace{5mm}  r^{2-d}\,\tilde S_d
\sum_{l=0}^{\infty}\left[\sigma_{ab}^{(1)}\,P^{-2}-1\right]^{-1}
{d-3+l\choose l}
\left\{2+\sigma_{ab}^{(3)}\;\left[\left(\frac{r}{R_+}\right)^{2\lambda
}
+\left(\frac{R_+}{r}\right)^{2\lambda}\right]\right\}\>.
\end{eqnarray}
Note that the second term in curly brackets is different from that in
(\ref{endi}).

Here we study the behavior for $r\rightarrow R_+$ with $b=O$,
$\sigma_{ab}^{(3)}=-1$. In this case the expression in curly
brackets in
(\ref{eddi}) tends to $-4\lambda^2\left[(R_+-r)/R_+\right]^2$ , and
the sum over $l$ can be expressed in terms of $\langle
T_{nn}\rangle$ in (\ref{tgau}). Using (\ref{tenn}), this leads to
\begin{equation}\label{p2li}
\langle \roarrow  \Phi^{\>2}(r)\rangle_{\rm \scriptscriptstyle CON}
\longrightarrow
\langle \roarrow  \Phi^{\>2}(r)\rangle_{\rm fc}\>\left[
1-\frac12\,\tilde
S_d^{-1}\;N^{-1}\;Y(\rho)\;\left(\frac{R_+-r}{R_+}\right)^d+
\ldots\right]
\end{equation}
for $r\rightarrow R_+$ and $b=O$. This is the counterpart of
(\ref{fidg}).
As in Eqs.\,(\ref{trfo})$-$(\ref{sham}), one may transform to an SPW
geometry and identify the leading coefficient $C_{\Phi^2}^{(O)}$ in
the short-distance expansion (SDE) about the planar wall with $b=O$.
Now exponents $-(d-2)$ and $d$ instead of $-1$
and $4$ appear
in Eqs.\,(\ref{trfo}) and (\ref{expw}), respectively, and
\begin{equation}\label{cphi}
C_{\Phi^2}^{(O)}=2^{d-1}/(N\tilde S_d)\>.
\end{equation}
This result, obtained here from the ${\rm S}_a
{\rm PW}_O$ geometry with $a=O$ or $SB$, is consistent with the
result for
${\rm S}_{\uparrow} {\rm PW}_O$ geometry for $d\rightarrow 4$,
discussed in the paragraph after Eq.\,(\ref{sham}). It is
also in agreement with
Refs.\, \cite{Krech94,EKD}, where the SDE of $\Phi^2$ about an $O$
surface has been applied to the profile in the
parallel-plate geometry and in the half--space
at $T>T_c$ and to correlation functions in the half--space at $T_c$. In
Ref.\,\cite{EKD} $C_{\Phi^2}^{(O)}$ for the Gaussian model was
called ${\cal B}$.

Eq.\,(\ref{eddi}) can also be used to evaluate the density
profiles for arbitrary fixed $0<r<R_+$ as $R_-\rightarrow 0$, and one
may check the predictions of the small-sphere expansion. This is
done in Appendix \ref{B2}.

Finally, we mention that for $\rho\rightarrow 1$
the profiles (\ref{endi}) of the energy density
reduce to the Gaussian results in the
parallel-plate geometry \cite{Krech94,EKD}. This may be
shown with arguments analogous to those near Eqs. (\ref{plim})
and (\ref{suli}).
\setcounter{equation}{0}

\section{Summary and concluding remarks}\label{5}
We have analyzed the Casimir interaction between two spherical
particles in a fluid at the critical point with boundary
conditions
$$(a,\,b) = (\uparrow,\,\uparrow), \>(\uparrow,\,\downarrow),
\>(\uparrow, O),\> (\uparrow, SB), \>(O,\,O), \>(O,\,SB),\>
(SB,\,SB)\>.$$
The Casimir interaction is completely determined by
a universal function ${\cal F} = {\cal F}_{a,b}$ of a single variable
$\kappa$ (see Eq.\ (\ref{kap2})),
which is a combination of the three lengths that characterize
the configuration of two spheres.

We have discussed the limiting behavior for spheres that
nearly touch $(\kappa\rightarrow 1)$ and spheres that are far apart
from each other $(\kappa \rightarrow \infty)$ and calculated the
complete functions
${\cal F}_{ab} (\kappa)$ for the case where the spatial dimension $d$
is near 4 \cite{foo13} (see Fig.\ 2).
{}From these results and corresponding information for
$d=2$ and $(a,\,b) = (\uparrow,\,\uparrow), (\uparrow,\,\downarrow),
(\uparrow, O), (O,\,O)$ \cite{B+E95},
one can make fairly reliable estimates of the scaling functions
${\cal F}_{ab}$ in $d=3$.

Since $\kappa$ and ${\cal F}$ are unchanged under
conformal coordinate transformations, it
is sufficient to consider a critical system bounded by two
concentric spheres. We have derived ${\cal F}(\kappa)$ by calculating
the average of the stress-energy tensor in this highly symmetric
geometry.

Besides the global free energy local
thermodynamic averages, e.g. profiles of the order parameter
and the energy density in the presence of two spherical particles,
are also of interest. These
can also be derived from the concentric geometry \cite{G+R}, and we
have presented explicit results for $d$ near 4 and for various
surface universality classes $a,\, b$.
We have also demonstrated that
the stress-tensor average determines the modification
of the density profile near the surface of one sphere due
to the other distant sphere \cite{FideG78,Cardy90}.

The Casimir interaction and the
density profiles for parallel plates are obtained
as a special case of the results for concentric spheres
when the radii
$R_-,R_+$ tend to infinity with fixed difference $R_+-R_-$ (compare
Fig.\ 1\,a). The other extreme $R_-\rightarrow 0$ with $R_+$
fixed is also of interest. It determines
the Casimir effect and the density profiles when one
sphere is much smaller than other lengths in the
problem. We have checked that the free energy contribution and
the density profiles arising from the presence of the small sphere are
consistent with a `small-sphere operator expansion'.
This method has been introduced in Ref.\ \cite{B+E95} and is also
explained in Sec.\ \ref{2C}
of this paper. The density profiles show still non-trivial scaling
forms in this limit (cf. Appendix B).

We have also found simple expressions \cite{foo35} for the interaction
between distant spheres in an infinite host fluid if the latter is
not at the critical point. Here both the host correlation length
$\xi$ and the distance $s$ between the spheres have to be much larger
than the radii of the spheres, but the ratio $\xi/s$ is unrestricted.
Similar expressions have been found for the interaction of a single
sphere with a distant planar boundary.

We conclude by discussing possible experimental observations of Casimir
interactions between spheres at the critical point. The case of
fluid mixtures at their
consolute point is of particular interest, since the attractive
interaction between two immersed equal spheres decays extremely slowly,
with a power law
\begin{equation}\label{5.1}
-k_BT_c\:{\cal A} \>(s_{12}/R_1)^{-1.04}
\end{equation}
in $d=3$ as the separation $s_{12}$ becomes much larger than the
radius $R_1$ of the spheres. The exponent follows from Eq.\
(\ref{fdis}), with $2\,x_\Phi \equiv 2\beta/\nu = 1.04$ in $d=3$
\cite{Guillou}.
In the only previous theoretical work we know on this topic, de Gennes
\cite{deG80} obtained an extremely good estimate $1$ for this exponent
by minimizing a phenomenological free energy functional.
The universal amplitude
${\cal A}=\left(A_{\uparrow}^{\Phi}\right)^2/B_{\Phi}$ is
estimated to be somewhat larger than $\sqrt{2}$  \cite{B+E95}.
The Casimir interaction (\ref{5.1}) should be compared with the van
der Waals interaction which decays much faster with an exponent $-6$
(non-retarded) or $-7$ (retarded). The prefactors are comparable. For
example, for fused silica spheres in water the van der Waals
prefactor is about $-k_B \cdot 10^3$ Kelvin \cite{Hunter}. We note
that for spheres in $^4$He at the $\lambda$--point
($XY$ model) the Casimir
exponent is $2x_{{\Phi}^2}\equiv 2(d-1/\nu) = 3.02$ for $d=3$.
Thus the decay is
faster than for fluid mixtures (see (\ref{5.1})).

Finally we mention
two classes of experiments one could conceivably carry out to
study the Casimir interaction:\vspace{2mm}\\
(i) Interaction between only two objects, for example a single sphere
near a  planar wall: By measuring the force on the sphere for varying
distance from the wall, one can in principle obtain the whole
scaling function ${\cal F}(\kappa)$.
Of course, the distance must not exceed
the correlation length of the nearly critical host binary fluid.
Correlation lengths of several $1000$\,\AA\,\,
are feasible, and one would like to have a sphere diameter of a
few hundred \AA.
Measurements of the interaction potential of isolated
pairs of spheres of somewhat larger size have been reported for a
charge stabilized colloid
\cite{PRL}, see also \cite{Ducker91}.
\vspace{2mm}\\
(ii) Aggregation phenomena in colloids: The Casimir interaction
between more than two spherical particles at large \cite{foo12}
separations is a sum of pair interactions,
which have the gravitation-like form (\ref{5.1})
for separations $s$ smaller than the correlation length $\xi$ of the
pure host and vanish for $s \gg \xi$.
On approaching the critical point
$\xi$ strongly exceeds the radii and mean separation of colloid
particles. This offers a possible mechanism for their aggregation
\cite{foo14}.
\vspace{4mm}

\noindent
{\bf Acknowledgements:} We thank W.\ Fenzl, M.\ L\"assig,
H.\ Wagner, and especially T.\ W.\ Burkhardt for helpful
comments and discussions. We also thank T.\ W.\ Burkhardt
for a critical reading of the manuscript.
U.\ R. would like to thank the
Deutsche Forschungsgemeinschaft for partial support through
Sonderforschungsbereich 237.\\[6mm]
\noindent
{\bf Note added in proof\,:} After this work was completed we realized that
the widely used non-retarded
Hamaker approximation $\delta E$ for the van der Waals
interaction of two spheres is also
{\it conformally invariant} (like the
critical Casimir energy $\delta F$
studied in this paper).
This follows
immediately from its representation
$-\left( A/ \pi^2\right)\:\int\:{\rm d}^3r_1\:{\rm d}^3r_2
\:|{\bf r}_1-{\bf r}_2|^{-6}$ with integrations over the two spheres
and allows to express $\delta E$
as a function of the invariant
quantity $\kappa$ defined in (1.4).
The explicit result [57] $\>\:\delta E\,(s_{12},\,R_1,\,R_2)=
-\left(A/6\right)\left\{ (\kappa-1)^{-1}+(\kappa+1)^{-1}
+{\rm ln}\left[(\kappa-1)/(\kappa+1)\right]\right\}$
may be used for a first comparison of critical Casimir and
van der Waals interaction. Note that for {\it close}
spheres ($\kappa$ near 1)
$\delta E$ has the {\it same} dependence
on $R_1,\, R_2,\, D$ as $\delta F$ and can be obtained from the right
hand side of (2.5) by replacing
$k_BT_c\,S_d\,\Delta_{ab}$ by $-A/3$ and setting $d=3$.

\setcounter{equation}{0}
\appendix
\section{ Free energy and stress-tensor average}\label{A}

An infinitesimal coordinate transformation
\begin{equation}\label{itra}
\widehat{{\rm \bf x}}={\rm \bf x} +{\rm\bf a}({\rm \bf x})
\end{equation}
changes a geometry $G$ to a geometry $\widehat G$. We consider two
examples:\\*[2mm]
(i) For the geometry of non-overlapping spheres in Fig.\ 1\,b where
${\rm \bf x}={\rm \bf r^{\,\prime}}$,
\begin{equation}\label{aeq1}
a_i({\rm \bf r})=\alpha\:\delta_{i,\bot}\:\Theta
\left(r_{\bot}^{\,\prime}\right)\>,
\end{equation}
with $\Theta$ the unit step
function, implies a rigid shift of the right sphere in the $\bot$
direction
and increases the distance $s_{12}$ between the spheres by the
amount
$\alpha$.\\*[2mm]
(ii) For the concentric
geometry in Fig.\ 1\,a where ${\rm
\bf x}={\rm \bf r}$, we consider the transformation (\ref{thet}).

The corresponding change in the singular part of the free energy is
given by \cite{CardyRev}
\begin{equation}\label{free}
F_{\widehat G}=F_G-k_BT_c\>\int{\rm d}^dx\>
\sum_{kl}\>\left(\partial a_k({\rm \bf
x})/\partial x_l
\right)\;\langle T_{kl}({\rm \bf x})\rangle_G
\end{equation}
to first order in $\alpha$. Here $T_{kl}$ is the stress tensor.
In case (i)
\begin{equation}\label{freg}
-\frac{\partial}{\partial s_{12}}\>\delta F(s_{12},\,R_1,\,R_2)/k_BT_c=
\int\:\mbox{d}^{d-1}r_{\parallel}^{\,\prime}\>\>
\langle T_{\!\bot\!\bot}(r_{\parallel}^{\,\prime},0)\rangle\>,
\end{equation}
with the average taken in the geometry of Fig.\ 1\,b. In case (ii)
\begin{equation}\label{part}
\partial a_k/\partial r_l=\alpha\left[
\delta_{kl}\Theta(r-r_0)+\frac{r_kr_l}{r}\,\delta(r-r_0)\right]\>.
\end{equation}
The result simplifies at the critical fixed point since the
trace of the stress tensor
vanishes for $R_-<r< R_+$ \cite{CardyRev};
a possible boundary contribution at $r=R_+$ drops out in the change
of $\delta F_{\scriptscriptstyle\rm CON}$ \cite{foo2}. Then only the
second term on the right hand side of (\ref{part}) contributes
and picks out
the radial
component $T_{nn}$ of the stress tensor.
This leads to
\begin{equation}\label{dfda}
\left.\frac{\rm d}{{\rm d}\alpha}\,F_{\scriptscriptstyle\rm
CON}\left(R_-/\left[(1+\alpha)R_+\right]\;\right)\>\right|_{\,\alpha=0}
= - k_BT_c\:
S_d\,r_0^d\;\langle T_{nn}(r_0)\rangle_{\scriptscriptstyle\rm
CON}
\end{equation}
to first order in $\alpha$, which implies Eq.\,(\ref{deef}).

Since both $\delta F(s_{12},R_1,R_2)$ and $\delta
F_{\scriptscriptstyle{CON}}(\rho)$ can be expressed
in terms of the same
function ${\cal F}(\kappa)$ (see Sec.\ \ref{1}), the ratio between
the left-hand sides of
Eqs.\ (\ref{freg}) and (\ref{deef}) is simply $-(\partial\kappa/\partial
s_{12})/
(\rho\,\mbox{d}\kappa(\rho)/\mbox{d}\rho)$, with $\kappa$ and
$\kappa(\rho)$
from (\ref{kap2}) and (\ref{kap3}), respectively.
It is instructive to verify from the transformation formula
(\ref{ttra}) for the stress-tensor average at the critical
fixed point that the right-hand sides of (\ref{freg}) and (\ref{deef})
lead to the same simple ratio. With $r_{\bot}^{\,\prime}=
{\rm \bf R}\cdot{\rm\bf r}^{\,\prime}=0$ and $r=R$ in Eq.\
(\ref{ttra}),
an easy $r_{\parallel}^{\,\prime}$-integration leads to the ratio
$1/(2R)$. This equals the above-mentioned ratio, since
\begin{equation}\label{aeq8}
2\,R\:\frac{s_{12}}{R_1\,R_2}=\frac12
\:\left(\rho^{-1}-\rho\right)
\end{equation}
as follows from writing $s_{12},\,R_1,\,R_2$ in terms of the
intersection points I, II, III, IV in (\ref{inte}).

Eq. (\ref{freg}) is not restricted
to the critical fixed point but applies
for arbitrary values $u_R$ of the renormalized $\Phi^4$ coupling
strength
(see Eq. (\ref{relu})). This can be used to show the universality
($u_R$-independence) of the singular part of the free energy in the
long-distance limit. For simplicity consider a situation where the
boundary conditions have their fixed-point form.
Then by naive dimensional analysis the
right hand side of (\ref{freg}) can be written as
\begin{equation}\label{aeq9}
\int\mbox{d}^{d-1}r_{\parallel}^{\,\prime}\>\>\langle T_{\!\bot\!\bot}
(r_{\parallel}^{\,\prime},\,0)\rangle=\frac{1}{s_{12}}
\>{\cal K}\left(R_1/s_{12}\,,\,R_2/s_{12}\,;\,\mu\,s_{12},\,u_R\right)
\>.
\end{equation}
In the `shift operator'
$\int\mbox{d}^{d-1}r_{\parallel}^{\,\prime}\>\>T_{\!\bot\!\bot}
(r_{\parallel}^{\,\prime},\,0)$ the improvement term ${\cal J}$
in (\ref{impr}) does not contribute, and the average (\ref{aeq9}) is
renormalized and
obeys a simple {\it homogeneous} renormalization group equation
\cite{Brown80}
\begin{equation}\label{aeq10}
\left[ \:\mu \,\frac{\partial}{\partial\,\mu}+\beta(u_R)\:
\frac{\partial}{\partial\,u_R} \right]
{\cal K}\left(R_1/s_{12},\,R_2/s_{12}\,;\,\mu\,s_{12},
\,u_R\right) = 0
\end{equation}
with the usual $\beta$-function \cite{Amit,ZinnJust}. Standard
arguments then imply that in the long-distance (infrared) limit
$\mu\,s_{12}\rightarrow \infty$
with $R_1/s_{12}$ and $R_2/s_{12}$ fixed
\begin{equation}\label{aeq11}
{\cal K}\left(R_1/s_{12},\,R_2/s_{12}\,;\,\mu\,s_{12},\,u_R\right)
\longrightarrow
{\cal K}\left(R_1/s_{12},\,R_2/s_{12}\,;\, 1,\,u_R^{*}\right)
\end{equation}
with $u_R^*$ from (\ref{usta}). This shows the $u_R$-independence of the
asymptotic critical Casimir force (\ref{freg}).

\section{ Density Profiles and Small-Sphere Expansion}\label{B}
Here we study the density profiles $\langle
\tilde\Psi\rangle_{\scriptscriptstyle \rm CON}\equiv \langle
\tilde\Psi\rangle_{a,b}$
in the
concentric geometry in the limit $R_-\rightarrow 0$ with
$\tilde r=r/R_+$
fixed at an arbitrary positive value $0<\tilde r<1$. We show that this
form is indeed in agreement with \cite{foo3a}, i.e.
\begin{equation}\label{pro1}
\langle \tilde\Psi({\rm \bf r})\rangle_{a,b}-\langle \tilde\Psi({\rm\bf
r})\rangle_{\rm fc}
\longrightarrow \frac{A_a^{\Psi}}{B_{\Psi}}\>
\left(R_-\right)^{x_{\Psi}}\Big[
\langle\Psi(0)\tilde\Psi({\rm\bf r})\rangle_{\rm fc}-
\langle\Psi(0)\rangle_{\rm fc}\langle\tilde\Psi({\rm\bf r})\rangle_{\rm fc}
\Big]\>,
\end{equation}
which follows from the small-sphere expansion (SSE) (\ref{tcon}),
(\ref{sms2}),
and (\ref{iden}). As before the full container (fc) has a surface with
universality class $b$. In (\ref{pro1}), $\Psi=\Phi$ if both
$A_a^{\Phi}$ and
the correlation in square brackets are non-vanishing, and
$\Psi=\Phi^2$ otherwise.
\subsection{Cases with broken symmetry
for $d\rightarrow 4$}\label{B1}
We start with the two cases
$(a\,b)=(\uparrow ,O)$ and $(\uparrow ,SB)$,
where only the inner sphere breaks the symmetry. Consider first
$\tilde\Psi=\Phi$.
Then $\Psi=\Phi$, and (\ref{pro1}) implies
\begin{mathletters}\label{beq3}
\begin{equation}\label{beq3a}
\langle\Phi({\rm\bf r})\rangle^{(0)}_{\uparrow ,b}=
\sqrt{\frac{6}{u}}\;\frac{v(r)}{r}
\end{equation}
with
\begin{equation}\label{beq3b}
v(r)\longrightarrow 2\sqrt{2}\,\rho\,q_{\mp}\>\quad{\rm where}\qquad
q_{\mp}=\tilde r^{-1}\mp \tilde r\>.
\end{equation}
\end{mathletters}\noindent
Here the upper sign holds for $b=O$ and the lower sign for
$b=SB$. To obtain (\ref{beq3}) we have used
the vanishing of $\langle\Phi\rangle_{\rm fc}$ for $b=O$ and $SB$
and the result for
$\langle\Phi(0)\Phi\rangle_{\rm fc}$ that
follows via a simple conformal
transformation \cite{B+E} from the half--space correlation
function, which for $d\rightarrow 4$ is
that of the Gaussian model with Dirichlet or
Neumann boundary conditions \cite{Diehl86}.
We have also used (\ref{leau}) and
(\ref{dokn}) to evaluate
$A_{\uparrow}^{\Phi}$ and $B_{\Phi}$, respectively.

The predictions (\ref{beq3}) of the SSE are consistent with the
solutions of the differential equation (\ref{dife}). For $b=O$ this
has the form
\begin{equation}\label{beq4}
{\rm ln}\frac{R_+}{r}=\sqrt{2}\int_0^{v}\frac{{\rm d}\widehat
v}{W(\widehat v)}\>.
\end{equation}
Note that $I$ is the same for $(a,b)=(\uparrow ,O)$ and
$(O,\uparrow)$ and equals $64\rho^2$ by (\ref{limi}) for our case
$\rho\rightarrow 0$. It follows from (\ref{beq3b}) that the
$\widehat v^4$-term in the function $W$ in (\ref{beq4}) is of order
$\rho^4$
and can be neglected, since $I$ and $2\widehat v^2$ are both of order
$\rho^2$. The integral in (\ref{beq4}) is then trivial and leads
to the result
(\ref{beq3b}) with the upper sign.

For $b=SB$, the solution follows from (\ref{beq4}) on replacing
the lower limit of integration by the positive $v$-value
\begin{equation}\label{beq5}
\left[\left(1+|I|\right)^{1/2}-1 \right]^{1/2}
\end{equation}
for which $W(v)$ vanishes. This guarantees that $dv/dr$ vanishes
at $r=R_+$ \cite{foo10a}.
For $\rho\rightarrow 0$, in which case $|I|\rightarrow
64\rho^2$ by Eq. (\ref{limi}), one may again neglect $\widehat v^4$
inside
$W(\widehat v)$, and the lower boundary of integration becomes
$(|I|/2)^{1/2}$. The integral is simple and leads
to the result (\ref{beq3b}) with the lower sign.

The leading contribution to the energy $\tilde \Psi=\Phi^2$
for $d\rightarrow 4$ is given
by \cite{foo11}
\begin{equation}\label{beq6}
\langle\Phi^2\rangle_{{\uparrow},b}^{(0)}=\left(
\langle\Phi\rangle_{{\uparrow},b}^{(0)}\right)^2
\end{equation}
and follows immediately from (\ref{beq3}). Since
$\Psi(0)=\Phi^2(0)$, the
right hand side of (\ref{pro1}) is, in leading order,
the square of its value in the previous cases,
and Eq.\ (\ref{pro1}) is again fulfilled.

Now we turn to the cases where the outer sphere breaks the symmetry,
e.g. $b=\uparrow$. For the order-parameter
profiles $\tilde \Psi=\Phi$ Eq.\ (\ref{pro1}) implies
\begin{mathletters}\label{beq7}
\begin{equation}\label{beq7a}
\langle\Phi({\rm \bf r})\rangle^{(0)}_{a,\uparrow}\longrightarrow
\sqrt{\frac{6}{u}}\>\,\frac{v^{({\rm fc})}_{\uparrow}(r)+
\delta v_{a,\uparrow}(r)}{r}
\end{equation}
with
\begin{equation}\label{beq7b}
v^{({\rm fc})}_{\uparrow}(r)=\frac{2\sqrt{2}}{q_-}
\end{equation}
and
\begin{equation}\label{beq7c}
\delta v_{a,\uparrow}(r)=-\frac{I}{16\sqrt{2}}\left\{
q_-+\frac{12}{q_-}\left[1-\frac{q_+}{q_-}\;{\rm ln}(\tilde
r^{-1})\right]
\right\}
\end{equation}
\end{mathletters}\noindent
with $I$ from (\ref{limi}) for the four cases $a=(\uparrow ,\,SB,\,
\downarrow ,\,O)$.
Here (\ref{beq7b}) represents the non-vanishing profile $\langle
\Phi\rangle_{\rm fc}$ in a full container (fc). It can be obtained by
means of a conformal transformation from the half--space profile
(\ref{leau}) as described
e.g. in Ref.\ \cite{B+E}. Eq. (\ref{beq7c}) respresents the
right hand side of
(\ref{pro1}). Here $\Psi(0)$ is $(\Phi,\,\Phi^2,\,\Phi,\,\Phi^2)$ in
the four cases
(see remark just below (\ref{dra4})). For $d\rightarrow 4$
\begin{mathletters}\label{beq8}
\begin{equation}\label{beq8a}
\langle\Phi(0)\Phi({\rm\bf r})\rangle_{\rm fc}-
\langle\Phi(0)\rangle_{\rm fc}\langle\Phi({\rm\bf r})\rangle_{\rm fc}
\longrightarrow K_{\rm fc}^{(G)}(r)
\end{equation}
is determined by Gaussian fluctuations $\varphi$ about the mean-field
profile
$\langle\Phi\rangle_{\rm fc}^{(0)}$. The quantity $K_{\rm fc}^{(G)}$
can be obtained from its counterpart in the half--space whose form is
known (see Eq. (4.96) in Ref.\ \cite{Jasnow}). The corresponding
correlation for $\Psi(0)=\Phi^2(0)$ follows from
\begin{equation}\label{beq8b}
\langle\Phi^2(0)\,\Phi({\rm\bf r})\rangle_{\rm fc}-
\langle\Phi^2(0)\,\rangle_{\rm fc}\langle\Phi({\rm\bf r})\rangle_{\rm
fc}
\longrightarrow 2\,
\langle\Phi(0)\rangle_{\rm fc}^{(0)}\;K_{\rm
fc}^{(G)}(r)\>.
\end{equation}
\end{mathletters}\noindent
Note the limits
\begin{mathletters}\label{beq9}
\begin{equation}\label{beq9a}
\delta v_{a,\uparrow}\rightarrow
-\frac{I}{16\sqrt{2}}\;\frac{R_+}{r}\quad,\qquad r \ll R_+
\end{equation}
and
\begin{equation}\label{beq9b}
\delta v_{a,\uparrow}\rightarrow - v_{\uparrow}^{({\rm fc})}\;
\frac{I}{40}\;\left(\frac{R_+-r}{R_+}\right)^4\quad ,\qquad
R_+-r\ll R_+
\end{equation}
\end{mathletters}\noindent
which follow from Eqs.\ (\ref{beq7}). Eq.\ (\ref{beq9b}) is
consistent with the expansion (\ref{fidg}) about the surface, which
applies for arbitrary $R_-/R_+$.

Again the prediction (\ref{beq7}) of the SSE may be checked by
comparing with the solution of the differential equation (\ref{dife}),
which is now given by Eq.\ (\ref{lnrr}). Since $v$ tends to the finite
positive value
$v_{\uparrow}^{({\rm fc})}$ in (\ref{beq7b}) as $I\rightarrow 0$, one
may expand the integrand in (\ref{lnrr}) with respect to $I$ and
invert the
function $r(v)$ to obtain $v=v(r)$, order by order in $I$.
In first order the result reads
\begin{equation}\label{beq10}
v(r)=v_{\uparrow}^{({\rm fc})}(r)\;\left\{
1-\frac{I}{\sqrt{2}}\;\frac{q_+}{q_-}\;\int_{v_{\uparrow}^{({\rm
fc})}(r)}^{\infty}\>{\rm d}\widehat v\>\left[W_0(\widehat
v)\right]^{-3}\>,
\right\}
\end{equation}
where $W_0$ is $W$ for $I=0$. The integral on the
right hand side is straightforward,
and one sees that Eq.\ (\ref{beq10}) indeed
implies the result (\ref{beq7}).
\subsection{Cases with symmetry preserved in the Gaussian
model}\label{B2}
Here we consider the $R_-\rightarrow 0$ behavior of the
left hand side of
(\ref{pro1}) for $\tilde\Psi=\Phi^2({\rm \bf r})$ and $a,\,b\in \{O,\,SB\}$.
The result
\begin{equation}\label{beq11}
\left[\langle\roarrow \Phi^2({\rm\bf r})\rangle_{ab}-
\langle\roarrow \Phi^2({\rm\bf  r})\rangle_{\rm fc}
\right]/N
\longrightarrow
\tilde S_d\,\sigma_{ab}^{(2)}\;(R_-)^{d-2}\>r^{-2(d-2)}\left[
1+\sigma_{ab}^{(3)}\tilde r^{d-2}\right]^2
\end{equation}
follows for the Gaussian model
from the $l=0$ term in (\ref{p2li}), using
$\sigma_{ab}^{(1)}\cdot\sigma_{ab}^{(3)}=\sigma_{ab}^{(2)}$.
Note that
$\sigma_{ab}^{(2)}$ only depends on $a$, while $\sigma_{ab}^{(3)}$ only
depends
on $b$. This is in agreement with the
right hand side of (\ref{pro1}).
It is easily verified that (\ref{beq11}) equals the
right hand side of
(\ref{pro1}): The leading operator is $\Psi=\Phi^2$, and
the $\Phi^2\cdot\Phi^2$ correlation function inside a full spherical
container with an $O$ or $SB$ surface, which appears on the
right hand side of (\ref{pro1}) in square brackets,
follows from its counterpart in the half--space
with Dirichlet or Neumann boundary conditions, respectively. The
prefactor $A_a^{\Phi^2}/B_{\Phi^2}$ in (\ref{pro1}) is obtained from
(\ref{cobe}) and (\ref{dokn}).

\newpage
\noindent{\Large\bf Figure Captions}\vspace{5mm}\\
{\bf Fig. 1:} Conformal mapping of concentric spheres with center at
the origin of ${\rm \bf r}$-space (Fig.\ 1\,a) onto spheres in ${\rm
\bf r}^{\,\>\prime}$-space with
centers on the line ${\rm \bf r}^{\,\prime}=r_{\bot}^{\,\prime}\,{\rm
\bf R}/R$ (Fig.\ 1\,b) as implied by Eq. (\ref{cotr}). The sphere
denoted by the broken line with radius $R$ is mapped onto the plane
$r_{\bot}^{\,\prime}=0$. The two points ${\rm \bf r} = -{\rm \bf
R}$ and ${\rm \bf r}= {\rm \bf R}$ of this sphere are mapped onto
${\rm \bf r}^{\,\>\prime}=0$ and ${\rm \bf r}^{\,\>\prime}=\infty$,
respectively.
The two spheres with radii $R_{-}$ and $R_{+}$ are mapped on spheres
with radii $R_1$ and $R_2$, respectively. The
corresponding intersection
points of the two spheres with the coordinate axes parallel ${\rm \bf
R}$ are denoted by I, II, III, IV.
For the case $R_-<R<R_+$ shown in Fig.\ 1\,a the images
${\rm \bf r}^{\,\prime}=2\,{\rm \bf R}$ and
${\rm \bf r}^{\,\prime}=-2\,{\rm \bf R}$ of ${\rm \bf r}=0$
and ${\rm \bf r}=\infty$ are located inside the
two spheres in Fig.\ 1\,b and are closer to the origin
than the centers of the spheres.\vspace{3mm}\\
{\bf Fig. 2:} Scaling functions ${\cal F}={\cal F}_{ab}$
in a critical Ising-type ($N=1$) system for spatial dimension
$d\rightarrow 4$. The function ${\cal F}_{O\,SB}$, which is not
shown, would nearly coincide with $-{\cal F}_{O\,O}=-
{\cal F}_{SB\,SB}$ in this plot. For the four symmetry-breaking
combinations $(a\,b)=(\uparrow\,\downarrow)$,
$(\uparrow\,\uparrow)$, $(\uparrow\,O)$, $(\uparrow\, SB)$, the
functions ${\cal F}$ behave as $1/\epsilon$ for $d\rightarrow 4$,
and we have plotted the finite limit $\tilde {\cal F}_{ab}=
\lim_{d\rightarrow 4} (4-d) \,{\cal F}_{ab}$.
\newpage
\def\epsfsize#1#2{0.75#1}
\hspace*{1cm}\epsfbox{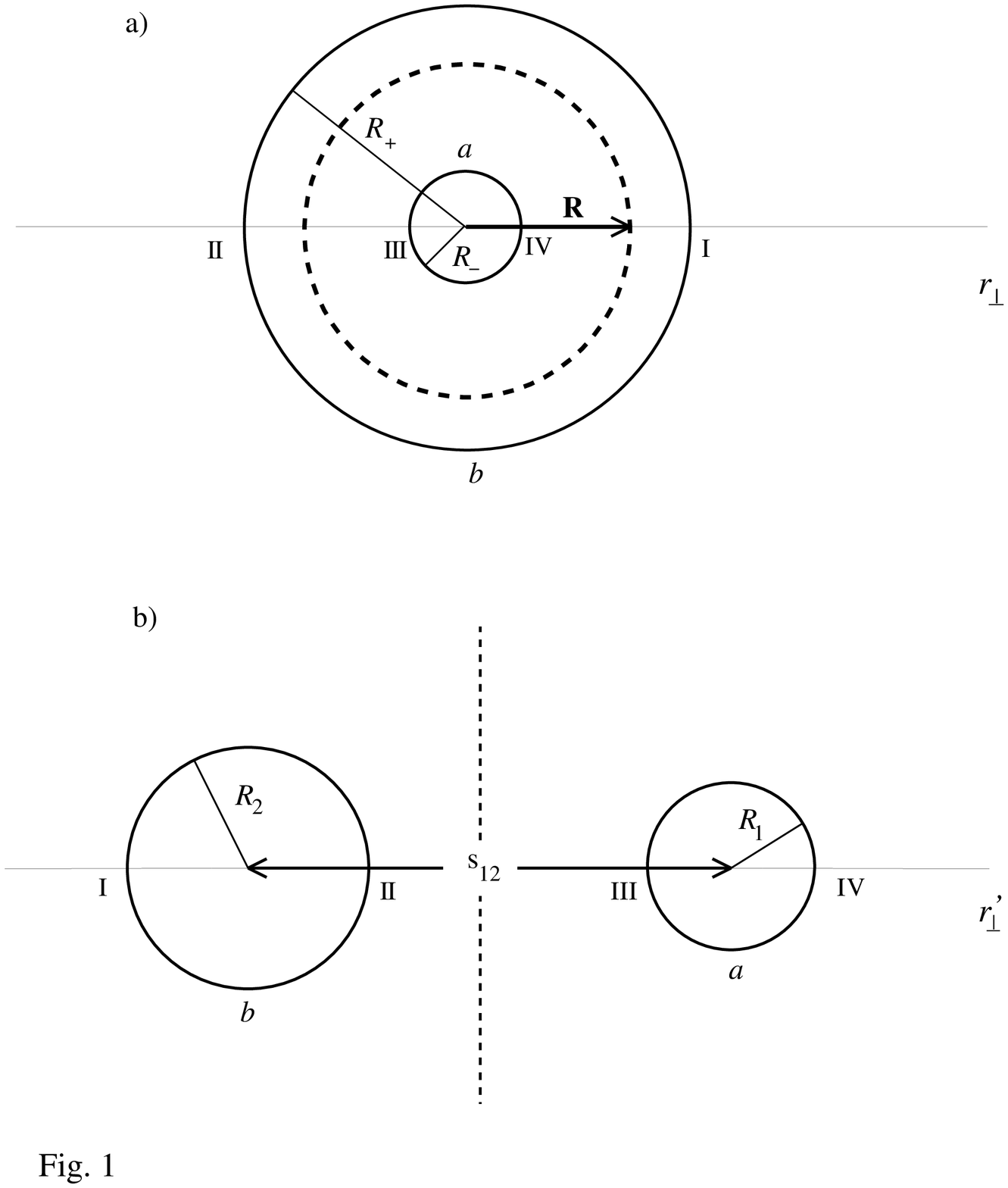}
\newpage
\def\epsfsize#1#2{0.75#1}
\hspace*{1cm}\epsfbox{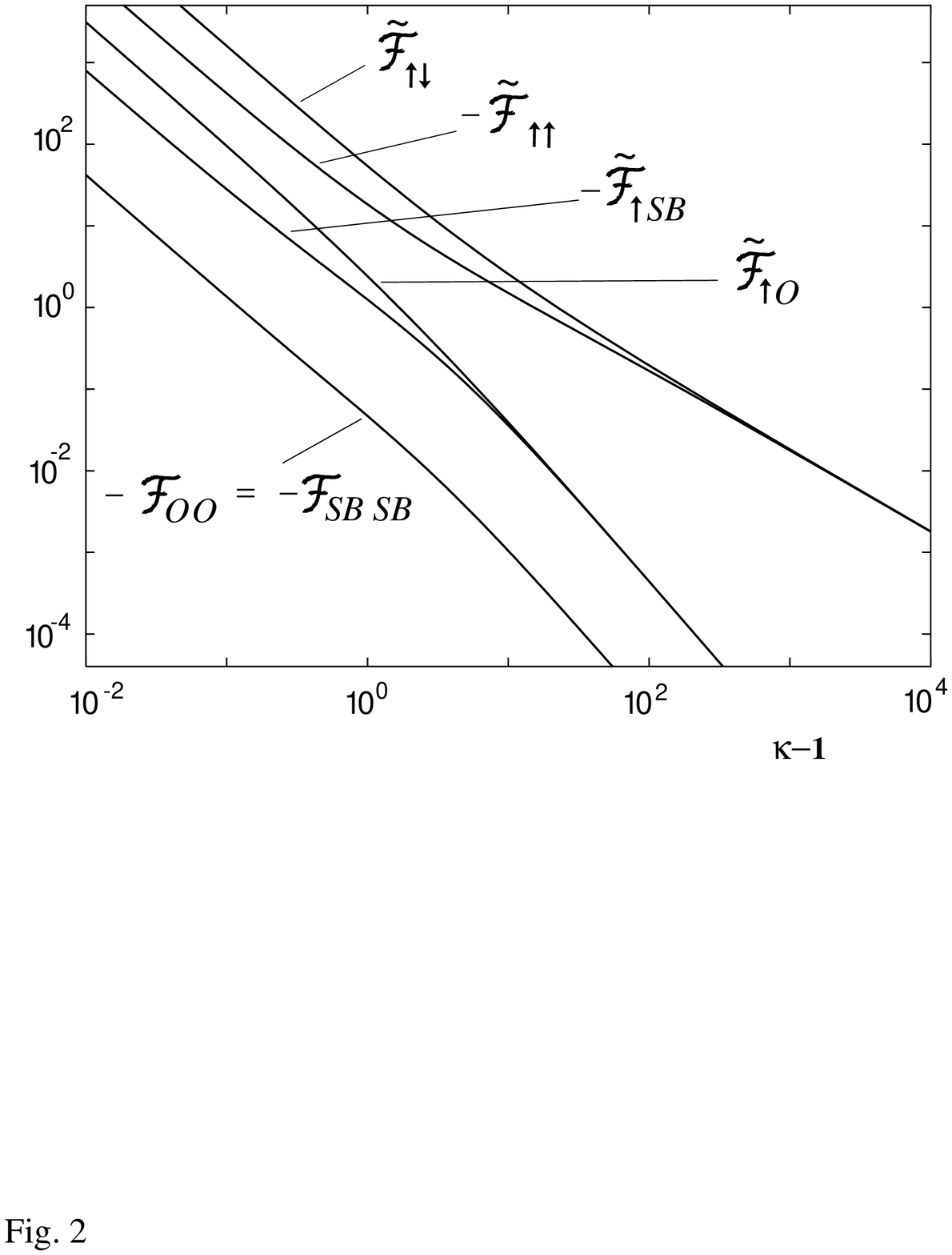}

\begin{thebibliography} {99}
\bibitem{foo1} Note that the static sphere in electromagnetism or
particle physics cannot be discussed in this way. The space-time
continuum where there is conformal invariance corresponds to a
cylinder and cannot be mapped onto the half--space in $d>2$.
\bibitem{foo3a} Here the tilde serves to distinguish
the primary operators $\tilde \Psi$ from those
which appear in the sum (\ref{sms2}). The latter
(which are special $\Xi$'s) are denoted by $\Psi$, compare
Eq.\ (\ref{iden}) below. Both $\tilde \Psi$ and $\Psi$
may be either the order parameter or the energy density.
\bibitem{foo4} We denote the normal--normal component of the stress
tensor by $T_{nn}$ in the concentric geometry of Fig.\ 1\,a, where the
normal is in the radial direction and by $T_{\!\bot\!\bot}$ in
geometries of the type depicted in Fig.\ 1\,b, where the normal is on
curved lines which are perpendicular to all the spherical ${\rm \bf
r}^{\,\prime}$-images of concentric ${\rm \bf r}$-spheres. In
particular the lines are perpendicular to the plane
$r_{\bot}^{\,\prime}=0$.
\bibitem{foo5} Here we consider situations where both the bulk and the
two surfaces are at critical fixed points of the renormalization
group. It is only in this case that dilatation and conformal
invariance hold on all length scales.
\bibitem{foo6} For our two-sphere geometries, it follows from
(\ref{tenn})
by means of the conformal transformation (\ref{cotr}) which itself is
regular near the boundaries.
\bibitem{foo8} Below we consider the Gaussian
model ($u=0$) for arbitrary spatial dimension $d>2$ and the
$\Phi^4$-model in dimensional regularization and within the
$\epsilon$-expansion. In both cases there is no $\Phi^2$ term in
${\cal L}$ at the bulk critical point.
\bibitem{foo8a} ${\rm d}v/{\rm d}r$ vanishes, however, at the $SB$
surface
(cf. Ref. \cite{G+R}).
\bibitem{foo8b} Here the integrand on the
right hand side of (\ref{lnrr}) is
integrated from zero, the value of $v$ at the $O$-surface, to $v$.
Close to the $O$-surface $v$ is small, and the integrand can be
expanded in powers of $\widehat v$.
\bibitem{foo13} While the complete functions ${\cal F}$ have
only been
obtained to leading order in the $\epsilon$-expansion,
most of the universal amplitudes that determine the limiting
behaviors are available to leading and next-to-leading order.
Corresponding results for the amplitude in (\ref{fdis}) for
$\kappa\rightarrow \infty$ can be inferred from
\cite{EiSt} and \cite{EKD} and are shown in Eqs.\
(20), (21) of \cite{B+E95}. The amplitudes in (\ref{plaw})
for $\kappa\rightarrow 1$ in the symmetry-preserving cases follow
from Ref.\ \cite{Krech91}.
\bibitem{foo12} If the separation is not much larger than the radius,
multiparticle interactions appear, which can be related to multispin
correlation functions in the Ising model \cite{B+E95}.
\bibitem{foo11} $\langle\Phi^2\rangle_{{\rm fc},b}$ is
non-vanishing for $b=O$ or
$SB$, but does not contain a contribution of order $1/u$.
\end{thebibliography}
\end{document}